\documentclass[11pt,a4paper,oneside]{JHEP3}
\usepackage{graphics,psfig,latexsym,amsthm,amsmath,amssymb}

\author{Carsten Kr\"uger\\ Institut f\"ur Theoretische Physik, Freie Universit\"at Berlin, 14195 Berlin\\ E-mail: ckrueger@physik.fu-berlin.de}
\title{Exact Operator Quantization of the Euclidean Black Hole CFT}
\abstract{We present an exact operator quantization of the Euclidean
Black Hole CFT using a recently established free field parametrization
of the fundamental fields of the classical theory \cite{IntegrationQuantumPara,MuellerWeigtIntegrPeriodic,MuellerComplSol,AnalSolMueller}. Quantizing
the map to free fields, we show that the resulting quantum fields
are causal and transform as covariant fields w.r.t. the Virasoro algebra.
We construct the reflection operator of the quantum theory and demonstrate its
unitarity. We furthermore discuss the W-algebra of the Euclidean Black Hole model. It turns out that unitarity of the 
reflection operator is a simple
consequence of the fact that certain representations of the W-algebra are
unitarily equivalent.}
\begin{document}
\newtheorem{de}{Definition:}
        \newcommand{\be}{\begin{eqnarray}}
        \newcommand{\ee}{\end{eqnarray}}
        \newcommand{\bemu}{\begin{multline}}
        \newcommand{\emu}{\end{multline}}
\tableofcontents
\section{Introduction}
The objective of the present article is to perform an exact quantization
 of the  conformal field theory (CFT)  classically defined by the action\\
\be\label{sigma}
S=\frac{1}{4\pi\alpha'}\int_{M} d\sigma dt \, \bigl (\partial_{\mu} r \partial^{\mu} r
+\tanh^{2} r\, \partial_{\mu} \theta \partial^{\mu} \theta \bigr )
\ee\\
which, given that $M$ has cylindrical topology, $M=\mathbb{R}\times S^{1}$,
 is the action of a so-called non-linear sigma model describing a
 closed string propagating in a target space with metric\\
\be\label{targetspace}
ds^{2}=dr^{2}+\tanh^{2} r\, d\theta^{2}
\ee\\
The target space  has the shape of a semi-infinite cigar if $\theta$ is an
 angular variable defined modulo $2\pi$.

It was realized in the early 90's by Witten \cite{witten} that this model may be formulated as a gauged
Wess-Zumino-Novikov-Witten (WZNW) theory \cite{wz,novi,witten2} based on the noncompact group $SL(2,\mathbb{R})$,
 the so-called $SL(2,\mathbb{R})/U(1)$ gauged WZNW model,  and has since
then attracted much attention  since the target space geometry (\ref{targetspace}) may be seen  to describe 
a 2d  Euclidean black hole. It is thus an important toy model of string theory in a two-dimensional
target space, and
Witten's discovery initiated quite some effort aiming at a better understanding of the Euclidean black hole
CFT. Let us just mention  the contribution of R. Dijkgraaf, E. Verlinde and H. Verlinde \cite{dijkgraafverlinde}, where
some interesting aspects of the dynamics of strings  propagating in the curved geometry (\ref{targetspace}) were
analyzed in the point-particle limit, which reduces  the problem to the quantization of the center of mass dynamics
of the string in an effective target space geometry. In this fashion, a scattering amplitude could be obtained, relating incoming and outgoing plane waves
on target space, and the spectrum of the theory could be determined.

Recently, the 2d black hole model has become important for the study of little string theories \cite{giveon1,giveon2,giveon3}. In this context exact
results on the 2- and 3-point function in the 2d black hole model have been
obtained by coset construction from the $H_{3}^{+}$-WZNW model \cite{opexH3,crossymH3,strucconst}. In this article
we will however follow an alternative approach. The classical theory defined by the action (\ref{sigma}) will be quantized
in a fashion that preserves conformal symmetry, and  our quantization approach will give rise to a conformal
quantum field theory defined on the world-sheet $M$ of the closed string.
We will use free field techniques very similar to those already applied to
Liouville theory \cite{LiouvilleRevisited}. Here, the classical Liouville
field is canonically mapped to a free field and then quantized by
quantizing the map to free fields. This method proves equivalently succesful for
the 2d black hole model, where the parametrization of the
classical Euclidean black hole fields in terms of free fields
has been established in \cite{IntegrationQuantumPara,MuellerWeigtIntegrPeriodic,MuellerComplSol,AnalSolMueller}. Taking into account that the free fields
parametrizing the black hole fields are asymptotic {\it in/out} fields
makes it possible to determine an exact expression for the scattering amplitude which in
the semi-classical limit reduces to the amplitude of \cite{dijkgraafverlinde}.
Another central result of this article refers to the well known W-algebra \cite{bakas} of
the Euclidean black hole model. It is demonstrated that representations
of the W-algebra to spins $j$ and $-j-1$ where $j\in -\frac{1}{2}+i\mathbb{R}$
are unitarily equivalent. To establish this fact we make use of the
representation of the W-currents in terms of free fields. It turns
out that unitarity of the reflection operator (or S-matrix) of the model is closely linked
to unitary equivalence of these representations.

\section{Structure of the classical solution}

\subsection*{Classical Dynamics}

Classically, the Euclidean Black Hole model may be defined
by gauge invariant reduction of the $SL(2,\mathbb{R})$ WZNW model.
The action obtained in this fashion reads
\be\label{action}
S=\frac{k}{4\pi}\int \limits_{M} d\sigma dt \,h^{\mu\nu}(\partial_{\mu}r\partial_{\nu}r+\tanh^{2} r \partial_{\mu} \theta \partial_{\nu} \theta)
\ee
with flat Minkowskian world-sheet metric $h_{\mu\nu}=\mathrm{diag}(+,-)$.
The world-sheet $M$ has cylindrical topology, i.e. $M=\mathbb{R}\times S^{1}$ corresponding to $\sigma\in [0,2\pi],\,\,-\infty<t<\infty$. The fields $r(\sigma,t),\theta(\sigma,t)$ map the world-sheet $M$ of a closed bosonic string
 into the curved target-space (\ref{targetspace}).
The fields $r,\theta$ obey the boundary conditions
\be
r(\sigma+2\pi,t)=r(\sigma,t), \qquad \theta(\sigma+2\pi,t)=\theta(\sigma,t)+2\pi\,w
\ee
We note that $\theta$ is not strictly periodic w.r.t. the space coordinate. Instead, we choose $\theta$ to be an angular variable defined modulo $2\pi$ only  in order to avoid a coordinate singularity at $r=0$. The target space then assumes the shape of a semi-infinite cigar. The integer $w$ is called the winding number and counts how often the string winds round the cigar.

In \cite{IntegrationQuantumPara,MuellerWeigtIntegrPeriodic}, the equations of motion derived from the action (\ref{action})
have been shown to be completely integrable, and a
parametrization of the on-shell physical fields $r,\theta$ in terms of
canonical free fields has been given. Let us briefly discuss
the structure of the classical solution. Introducing complex
Kruskal coordinates,
\be
u(\sigma,t)=\sinh r(\sigma,t)e^{i\theta(\sigma,t)}, \qquad \bar{u}(\sigma,t)=\sinh r(\sigma,t)e^{-i\theta(\sigma,t)}
\ee
the equations of motion assume the form
\be
\partial_{+}\partial_{-}u=\bar{u}\frac{\partial_{+}u\partial_{-}u}{1+u\bar{u}},\qquad \partial_{+}\partial_{-}\bar{u}=u\frac{\partial_{+}\bar{u}\partial_{-}\bar{u}}{1+u\bar{u}}
\ee
Here we have introduced light-cone coordinates according to
\be
\begin{array}{ll}
x_{+}=t+\sigma,\quad & x_{-}=t-\sigma \\
\partial_{+}=\frac{1}{2}(\partial_{t}+\partial_{\sigma}), \quad &
\partial_{-}=\frac{1}{2}(\partial_{t}-\partial_{\sigma})
\end{array}
\ee
The on-shell physical fields $u,\bar{u}$ may then be parametrized by canonical free fields
\be
\Phi_{1}(\sigma,t)=\phi_{1}(x_{+})+\bar{\phi}_{1}(x_{-}), \qquad \Phi_{2}(\sigma,t)=\phi_{2}(x_{+})+\bar{\phi}_{2}(x_{-})
\ee
It will be useful to recall here the mode expansions for the free fields. For the (anti)chiral components of $\Phi_{1}$ we have
\be
\phi_{1}(x_{+})&=&\frac{q_{1}}{2}+p_{1}x_{+}+i\sum_{n \not= 0} \frac{a_n^{(1)}}{n}e^{-inx_{+}}\\ \bar{\phi}_{1}(x_{-})&=&\frac{q_{1}}{2}+p_{1}x_{-}+i\sum_{n \not= 0} \frac{b_n^{(1)}}{n}e^{-inx_{-}}
\ee
Unlike $\Phi_{1}$, the free boson $\Phi_{2}$ is chosen to be
compactified on a circle of radius $R=\sqrt{k}$ which gives
a slightly modified mode expansion for its (anti)chiral components,
\be
\phi_{2}(x_{+})&=&\frac{q_{2}}{2}+(p_{2}+\frac{\sqrt{k}}{2}w)x_{+}+i\sum_{n \not= 0} \frac{a_n^{(2)}}{n}e^{-inx_{+}}\\
\bar{\phi}_{2}(x_{-})&=&\frac{q_{2}}{2}+(p_{2}-\frac{\sqrt{k}}{2}w)x_{-}+i\sum_{n \not= 0} \frac{b_n^{(2)}}{n}e^{-inx_{-}}
\ee
with integer winding number $w$.
A canonical map from the free fields to the physical fields $u,\bar{u}$
is then given by
\be\label{para}
u[\Phi_1,\Phi_2]=f_{1}(x_{+})\bar{f}_{1}(x_{-})-f_{2}(x_{+})\bar{f}_{2}(x_{-})
\ee
with building blocks
\be
\begin{array}{ll} 
f_{1}(x_{+})=e^{\frac{1}{\sqrt{k}}(\phi_{1}(x_{+})+i\phi_{2}(x_{+}))},\quad &
f_{2}(x_{+})=A(x_{+})f_1 (x_{+}) \\
\bar{f}_{1}(x_{-})=e^{\frac{1}{\sqrt{k}}(\bar{\phi}_{1}(x_{-})+i\bar{\phi}_2 (x_{-}))},\quad &
\bar{f}_{2}(x_{-})=\bar{A}(x_{-})\bar{f}_{1}(x_{-})
\end{array}
\ee
The screening charges $A(x_{+}),\bar{A}(x_{-})$ are defined as follows. Introducing the quantities
\be
V(x_{+})&=&\frac{1}{\sqrt{k}}\bigl (\partial\phi_{1}(x_{+})+i\partial\phi_{2}(x_{+})\bigr )e^{-\frac{2}{\sqrt{k}}\phi_{1}(x_{+})}\\
\bar{V}(x_{-})&=&\frac{1}{\sqrt{k}}\bigl (\partial\bar{\phi}_{1}(x_{-})+i\partial\bar{\phi}_{2}(x_{-})\bigr )e^{-\frac{2}{\sqrt{k}}\bar{\phi}_{1}(x_{-})}
\ee
the screening charges $A,\bar{A}$ are defined as solutions
of the differential equations
\be
\partial A(x_{+})=V(x_{+})\, , \qquad \partial \bar{A}(x_{-})=\bar{V}(x_{-})
\ee
The solutions to these equations are unique if
 we require the monodromy of the quantities $V,\bar{V}$
to be preserved under integration, i.e. with
\be
V(x_{+}+2\pi)=e^{-\frac{4\pi}{\sqrt{k}}p_{1}}V(x_{+})
\ee
we require the screening charges to have the same monodromy. One may
check that the following expression for $A(x_{+})$ solves this
problem,
\begin{multline}
A(x_{+})\\=-\frac{e^{\frac{2\pi}{\sqrt{k}}p_{1}}}{2\sqrt{k}\sinh \frac{2\pi}{\sqrt{k}}p_{1}}\int \limits_{0}^{2\pi}d\varphi\, \bigl (\partial\phi_{1}(x_{+}+\varphi)+i\partial\phi_{2}(x_{+}+\varphi)\bigr )e^{-\frac{2}{\sqrt{k}}\phi_{1}(x_{+}+\varphi)}
\end{multline}
The corresponding  expression for $\bar{A}$ may be obtained by
obvious replacements.

To motivate the discussion of the quantum mechanical reflection operator of the model to be discussed
in subsection 3.9, it is essential to point out that
the free fields parametrizing $u$ are $in/out$ fields defined by
\be
\begin{array}{ll}\label{inout}
r(\sigma,t) \overset{t\rightarrow -\infty}{\sim} \frac{1}{\sqrt{k}}\Phi_{1}^{in}(\sigma,t), \quad & \theta(\sigma,t) \overset{t\rightarrow -\infty}{\sim}\frac{1}{\sqrt{k}}\Phi_{2}^{in}(\sigma,t) \\ & \\
r(\sigma,t) \overset{t\rightarrow +\infty}{\sim} \frac{1}{\sqrt{k}}\Phi_{1}^{out}(\sigma,t), \quad & \theta(\sigma,t) \overset{t\rightarrow +\infty}{\sim}\frac{1}{\sqrt{k}}\Phi_{2}^{out}(\sigma,t)
\end{array}
\ee
As the string propagates freely in the asymptotically flat region $r\rightarrow \infty$,
the in/out fields defined in this fashion are indeed free fields as long as we consider
solutions that satisfy $r_{0}(t)\rightarrow \infty$ for large positive and negative
times, where $r_{0}(t)$ represents the zero mode of field $r(\sigma,t)$.
Let us note that the free in-field $\Phi_{1}^{in}$ travelling towards the tip of
the cigar at $r=0$ has negative momentum zero-mode $p_{1}^{in}<0$,
while the outgoing field $\Phi_{1}^{out}$ has
 momentum zero-mode $p_{1}^{out}=-p_{1}^{in}>0$. It follows that the field $u$ has the
asymptotic  behaviour
\be\label{asymp}
u(\sigma,t)&\overset{t \rightarrow -\infty}{\sim}&e^{r+it}=e^{\frac{1}{\sqrt{k}}(\Phi_{1}^{in}+i\Phi_{2}^{in})}\\
u(\sigma,t)&\overset{t \rightarrow +\infty}{\sim}&e^{r+it}=e^{\frac{1}{\sqrt{k}}(\Phi_{1}^{out}+i\Phi_{2}^{out})}
\ee
Now suppose we were given some solution $u(\sigma,t)$ of the dynamical equations  that asymptotically behaves according to (\ref{asymp}). The assertion is that the full interacting field $u(\sigma,t)$ for all times $t$ is given by
\be
u(\sigma,t)=u[\Phi_{1}^{in},\Phi_{2}^{in}]=u[\Phi_{1}^{out},\Phi_{2}^{out}]
\ee
But this is easily proven by considering the asymptotic behaviour
of $u[\Phi_{1}^{in/out},\Phi_{2}^{in/out}]$. Taking properly into account
the dependence of the free-field exponentials on the momentum zero-mode of $\Phi_{1}^{in/out}$ we find
\be
u[\Phi_{1}^{in},\Phi_{2}^{in}]&\overset{t \rightarrow -\infty}{\sim}&e^{\frac{1}{\sqrt{k}}(\Phi_{1}^{in}+i\Phi_{2}^{in})}
\ee
and similarly
\be\label{outplusasymp}
u[\Phi_{1}^{out},\Phi_{2}^{out}]&\overset{t \rightarrow +\infty}{\sim}&e^{\frac{1}{\sqrt{k}}(\Phi_{1}^{out}+i\Phi_{2}^{out})}
\ee
The term containing the screening charges is supressed in both cases. Comparison with (\ref{asymp}) leads us to conclude that
the free fields parametrizing $u$ are asymptotic $in/out$ fields, depending on the sign of momentum zero-mode of $\Phi_{1}$. Finally, let us determine the
 classical ``S-matrix'' of our model, i.e. we would like to
find some map between the $in$- and the $out$-fields. The key
to the construction of this map, which we call $S$, is the
fact that the $in$ and the corresponding $out$-fields
parametrize the same solution $u(\sigma,t)$. For large positive times consider the
asymptotics
\be
u[\Phi_{1}^{in},\Phi_{2}^{in}]&\overset{t \rightarrow +\infty}{\sim}&=-A(x_{+})\bar{A}(x_{-})e^{\frac{1}{\sqrt{k}}(\Phi_{1}^{in}+i\Phi_{2}^{in})}
\ee
Comparison with (\ref{outplusasymp}) yields the following relation
between the $in$ and the corresponding $out$-field:
\be
e^{\frac{1}{\sqrt{k}}(\Phi_{1}^{out}+i\Phi_{2}^{out})}=-A(x_{+})\bar{A}(x_{-})e^{\frac{1}{\sqrt{k}}(\Phi_{1}^{in}+i\Phi_{2}^{in})}
\ee
Similarly, we may derive the inverse relation
\be
e^{\frac{1}{\sqrt{k}}(\Phi_{1}^{in}+i\Phi_{2}^{in})}=-A(x_{+})\bar{A}(x_{-})e^{\frac{1}{\sqrt{k}}(\Phi_{1}^{out}+i\Phi_{2}^{out})}
\ee
What are the conclusions we can draw concerning
the phase space of the Euclidean black hole model ? To answer this question, introduce a splitting of the free field phase space $\mathcal{P}^{F}$
according to $\mathcal{P}^{F}=\mathcal{P}^{F}_{+}\oplus \mathcal{P}^{F}_{-}$,
where $\mathcal{P}^{F}_{\pm}$ corresponds to the subspace of free field configurations with positive resp. negative momentum zero-mode $p_{1}$.
Then,  $S$ is a map $S: \mathcal{P}^{F}_{\pm}\rightarrow \mathcal{P}^{F}_{\mp}$
by means of which an equivalence relation $\sim$ may be introduced
on $\mathcal{P}^{F}$ identifying free field configurations which are
images of each other under $S$. As equivalent free field configurations
parametrize the same solution $u(\sigma,t)$ of the dynamical equations,
the phase space of the Euclidean black hole model is isomorphic to $P^{F}/\sim$. In the following, our aim will be to carry over these considerations to
the quantum mechanical model.

\subsection*{Conserved classical currents}
As a classical field theory, the $SL(2,\mathbb{R})/U(1)$ model is
conformally invariant. Conformal transformations are generated
by the (anti)chiral components of the energy momentum tensor
\be\label{uem}
T(x_{+})=k\frac{\partial_{+}\bar{u}\partial_{+}u}{1+u\bar{u}}\, , \quad
\bar{T}(x_{-})=k\frac{\partial_{-}\bar{u}\partial_{-}u}{1+u\bar{u}}
\ee
Interestingly, the Fourier modes
$l_{n}=\frac{1}{2\pi}\int \limits_{0}^{2\pi} d\sigma T(\sigma)e^{in\sigma}$ of the energy momentum tensor
satisfy a Poisson counterpart of the Virasoro algebra
with vanishing central charge,
\be
\{l_{n},l_{m}\}=i(n-m)l_{n+m}
\ee
Substituting the free field
parametrization into (\ref{uem}) transforms the components of
the energy momentum tensor into that of a free theory,
\be
T^F(x_{+})=\big (\partial\phi_{1}(x_{+}) \big)^{2}+\big (\partial\phi_{1}(x_{+}) \big)^{2}\, , \quad 
\bar{T^F}(x_{-})=\big (\partial_{-}\bar{\phi}_{1}(x_{-}) \big)^{2}+\big (\partial_{-}\bar{\phi}_{1}(x_{-}) \big)^{2}
\ee
As the map to free fields preserves the symplectic structure of the theory, the modes
of $T^{F}$ satisfy the same Poisson Virasoro algebra as that of $T$.

Besides the energy momentum tensor, another set of conserved currents can be found, as it is well known that upon
 reduction to the coset $SL(2,\mathbb{R})/U(1)$
the classical Kac-Moody currents of the ungauged $SL(2,\mathbb{R})$ WZNW-model
reduce to parafermionic coset currents. As argued for in \cite
{IntegrationQuantumPara}, 
these parafermionic currents are given in terms of free fields as
\be
\psi^{\pm}=i(\partial \phi_{1}(x_{+})\pm i \partial \phi_{2}(x_{+}))e^{\pm \frac{2i}{\sqrt{k}} \phi_{2}(x_{+})}
\ee
It will turn out that a natural starting point for
quantization is  the quantization of these currents as
 it is known that the chiral algebra of the quantum mechanical
 $SL(2,\mathbb{R})/U(1)$
model is generated by fusion of the quantum parafermions.

\section{Quantization}

\subsection*{Quantum parafermions and energy momentum tensor}

Let us introduce (anti)holomorphic Euclidean free fields $\phi_{1}(z),\phi_{2}(z),\bar{\phi}_{1}(\bar{z}),\bar{\phi}_{2}(\bar{z})$
with short-distance behaviour
\be
\phi_{k}(z)\phi_{l}(w)=-\frac{1}{2}\delta_{kl}\ln (z-w)\, , \qquad
\bar{\phi}_k (\bar{z})\bar{\phi}_l (\bar{w})=-\frac{1}{2}\delta_{kl} \ln (\bar{z}-\bar{w})
\ee
Then the quantum analoga of the classical parafermions turn out to be
\be\label{paraferm}
\psi^{\pm}(z)=i:(\eta \partial \phi_{1}(z)\pm i \partial \phi_{2}(z))e^{\pm
\frac{2i}{\sqrt{k}}\phi_{2}(z)}:
\ee
The corresponding antiholomorphic currents are obtained by obvious
 replacements, we will however restrict the discussion here to the
holomorphic copy of the symmetry algebra.
 The deformation parameter $\eta$
depends on the level $k$ only and is given by
\be
\eta=\sqrt{\frac{k-2}{k}}
\ee
It turns out that the introduction of $\eta$ in (\ref{paraferm}) is necessary for a consistent quantization and
leads to the
canonical form of the parafermion OPE as first introduced in \cite{zamo2}.
Indeed, a standard calculation using Wick's contraction theorem for free fields
yields
\be\label{ope}
\psi^{+}(z)\psi^{-}(w)=(z-w)^{-2\Delta_{\psi}} \bigg \{ 1+\frac{2\Delta_{\psi}}{c}(z-w)^{2}\mathsf{T}(w)+\mathcal{O}((z-w)^{3})\bigg \}
\ee
where $c$ and $\Delta_{\psi}$ are the central charge resp. the conformal weight of the parafermions, see below.
The spin 2 current on the r.h.s. can be identified
with the energy momentum tensor of the quantum theory. It has the free field parametrization
\be
\mathsf{T}(z)=-:(\partial \phi_{1}(z))^{2}:-:(\partial \phi_{2}(z))^{2}:-b\,\partial^{2}\phi_{1}(z)
\ee
where we introduced the parameter
\be
b=\frac{1}{\sqrt{k-2}}
\ee
A background charge $Q=-b$ that was absent classically, enters the quantum theory. Conformal symmetry in the quantum theory is
generated by the Fourier-Laurent modes of $\mathsf{T}(z)$ which
 satisfy
a Virasoro algebra
\be
[\mathsf{L}_{n},\mathsf{L}_{m}]=(n-m)\mathsf{L}_{n+m}+\frac{c}{12}n(n^{2}-1)\delta_{n+m},0
\ee
The central charge is found to be given in terms of the level $k$ as
\be
c=\frac{3k}{k-2}-1
\ee
This precisely coincides with what was found by Witten for the
black hole coset model, see \cite{witten}.

Having identified the energy momentum tensor of the quantum theory, we find
that the parafermionic currents obey the following commutation relations
with the modes of the energy momentum tensor,
\be
[\mathsf{L}_{n},\psi^{\pm}(z)]=z^{n}(z\partial+\Delta_{\psi}(n+1))\psi(z)
\ee
which implies that they are primary with conformal dimension $\Delta_{\psi}=1+\frac{1}{k}$.
\subsection*{The W-algebra of the Euclidean black hole model}
A remark on the W-algebra of the Euclidean black hole CFT is in order, as
it will turn out that the quantum counterpart of the map $S : \mathcal{P}^{F}_{\pm}
\rightarrow \mathcal{P}^{F}_{\mp}$ intertwines certain representations of this
algebra. The currents of this algebra appear on the r.h.s. of the OPE (\ref{ope}). Indeed, there appears an infinite set of integer 
spin currents $\mathsf{W}_{s}(z)$ which constitute a closed nonlinear
operator algebra.  They furthermore transform as primary fields of the
Virasoro algebra,
\be
[\mathsf{L}_{n},\mathsf{W}_{s}(z)]=z^n (z\partial +s(n+1))\mathsf{W}_{s}(z)
\ee
For example, including terms of order $(z-w)^3$ inside the
curly bracket on the r.h.s. of (\ref{ope}), we find the OPE
\begin{multline}
\label{OPE3}
\psi^{+}(z)\psi^{-}(w)=\\=(z-w)^{-2\Delta_{\psi}}\bigg \{1+\frac{2\Delta_{\psi}}{c}(z-w)^{2}\,\,\bigg (\mathsf{T}(w)+\frac{1}{2}(z-w)\partial \mathsf{T}(w)\bigg )-(z-w)^{3}\,\,\mathsf{W}_{3}(w)+\mathcal{O}((z-w)^{4}) \bigg \}
\end{multline}
where the spin 3 current $\mathsf{W}_{3}(z)$ is given in terms
of free fields as
\begin{multline}
\label{omega3}
\mathsf{W}_{3}(z)=\alpha\,\,:(\partial\phi_{2}(z))^{3}:+\beta\,\, \partial^{3}\phi_{2}(z)+\gamma\,\, :(\partial\phi_{1}(z))^{2}\partial\phi_{2}(z): \\+\delta\,\,
\partial^{2}\phi_{2}(z)\partial\phi_{1}(z)+\varepsilon\,\, \partial^{2}\phi_{1}(z)\partial\phi_{2}(z)
\end{multline}
with coefficients given as functions of $k$ by
\be
\label{coeff}
&\alpha=\frac{2i}{3}\frac{3k-4}{k\sqrt{k}} \qquad  \beta=\frac{1}{6}\frac{i}{\sqrt{k}} \qquad   \gamma=2i \frac{k-2}{k\sqrt{k}} \nonumber \\ &\delta=i\sqrt{\frac{k-2}{k}} \qquad    \epsilon=-i\frac{k-2}{k}\sqrt{\frac{k-2}{k}}
\ee
For more details on the W-algebra of the Euclidean black hole model, see
for example \cite{bakas}, let us just add one important
property of the currents of this algebra. It turns out that the structure of the OPE implies that the currents
$\mathsf{W}_{s}(z)$ transform under parity $\phi_{2} \rightarrow -\phi_{2}$ as
\be
\mathsf{W}_{s}[\phi_{1},-\phi_2]=(-)^s \mathsf{W}_{s}[\phi_{1},\phi_2]
\ee
We will see that using this relation one may show
that certain representations of the W-algebra are unitarily equivalent,
which implies unitarity of the quantum mechanical reflection operator.

\subsection*{Free bosonic quantum fields in Minkowskian $\mathbb{R}\times S^{1}$}

As the quantization of the Euclidean black hole fields $u,\bar{u}$
will be done throughout in the Minkowskian formulation of CFT, we shall
briefly discuss the quantization of the free bosons $\Phi_{1},\Phi_{2}$
on the Minkowskian cylinder $\mathbb{R}\times S^{1}$. The chiral field $\phi_{1}(x_{+})$ and its antichiral counterpart
corresponding to the holomorphic resp. antiholomorphic
 Euclidean fields introduced
in the previous subsection have the well-known
mode expansion 
\be
\phi_{1}(x_{+})&=&\mathsf{q}_{1}+\mathsf{p}_{1}x_{+}+i\sum_{n \not= 0}\frac{\mathsf{a}_{n}^{(1)}}{n}e^{-inx_{+}}\\
\bar{\phi}_{1}(x_{-})&=&\mathsf{q}_{1}+\mathsf{p}_{1}x_{-}+i\sum_{n \not= 0}\frac{\mathsf{b}_{n}^{(1)}}{n}e^{-inx_{-}}
\ee
For the chiral and antichiral components of compactified boson $\Phi_{2}$,
we introduce  commuting copies of the zero-mode algebra for
the left and right-moving sectors. The mode
expansion then reads
\be
\phi_{2}(x_{+})&=&\mathsf{q}_{2}^{L}+\mathsf{p}_{2}^{L}x_{+}+i\sum_{n \not= 0}\frac{\mathsf{a}_{n}^{(2)}}{n}e^{-inx_{+}}\\
\bar{\phi}_{2}(x_{-})&=&\mathsf{q}_{2}^{R}+\mathsf{p}_{2}^{R}x_{-}+i\sum_{n \not= 0}\frac{\mathsf{b}_{n}^{(1)}}{n}e^{-inx_{-}}
\ee
The nonvanishing commutators for the zero-modes are
\be
[\mathsf{q}_{1},\mathsf{p}_{1}]=\frac{i}{2}, \qquad [\mathsf{q}_{2}^{L},\mathsf{p}_{2}^{L}]=\frac{i}{2}, \qquad  [\mathsf{q}_{2}^{R},\mathsf{p}_{2}^{R}]=\frac{i}{2}
\ee
and for the oscillators
\be
[\mathsf{a}_{k}^{(i)},\mathsf{a}_{l}^{(j)}]=\frac{k}{2}\delta_{ij}\delta_{n+m,0}\quad, \qquad [\mathsf{b}_{k}^{(i)},\mathsf{b}_{l}^{(j)}]=\frac{k}{2}\delta_{ij}\delta_{n+m,0}
\ee
The commutation relations have to be supplemented by hermiticity relations,
which for the oscillators and zero modes read
\be
\begin{array}{cc}
\mathsf{q}_{1}^{\dagger}=\mathsf{q}_{1}, \quad &
(\mathsf{q}_{2}^{L/R})^{\dagger}=\mathsf{q}_{2}^{L/R}\\
\mathsf{p}_{1}^{\dagger}=\mathsf{p}_{1}, \quad &
(\mathsf{p}_{2}^{L/R})^{\dagger}=\mathsf{p}_{2}^{L/R}\\
(\mathsf{a}^{(k)}_{n})^{\dagger}=\mathsf{a}^{(k)}_{-n}\, ,\quad &
(\mathsf{b}^{(k)}_{n})^{\dagger}=\mathsf{b}^{(k)}_{-n}
\end{array}
\ee
The Hilbert space on which the oscillators and zero-modes
take their action admits a decomposition into
Fock modules which diagonalize the action of the zero modes 
 $\mathsf{p}_{1},\mathsf{p}_{2}^{L/R}$. 
The compactification of 
free boson $\Phi_{2}$ implies that $\mathsf{p}_{2}^{L/R}$
have discrete spectra
\be
\mathrm{spec}\, \mathsf{p}_{2}^{L}&=&\{\frac{1}{2\sqrt{k}}(m+nk)|m,n\in \mathbb{Z}\}\\
\mathrm{spec}\, \mathsf{p}_{2}^{R}&=&\{\frac{1}{2\sqrt{k}}(m-nk)|m,n \in \mathbb{Z}\}
\ee
whereas the momentum zero mode $p_{1}$ of uncompactified free boson $\Phi_{1}$ has real continuous spectrum. We note that $m$ resp. $n$ correspond to the momentum resp.
winding number in the compactified $\theta$-direction in asymptotic regions
of the cigar and correspond to the eigenvalues of the operators
\be
\mathsf{m}=\sqrt{k}(\mathsf{p}_{2}^{L}+\mathsf{p}_{2}^{R})\,, \quad
\mathsf{n}=\frac{1}{\sqrt{k}}(\mathsf{p}_{2}^{L}-\mathsf{p}_{2}^{R})
\ee
The decomposition of the Hilbert space $\mathcal{H}^{F}$ into
charged Fock modules then reads
\be
\mathcal{H}^{F}=\bigoplus_{m,n \in \mathbb{Z}} \int^{\oplus}_{j=-1/2+i\rho}
dj\, \mathcal{F}^{j}_{mn}\otimes \mathcal{F}^{j}_{mn}
\ee
Here, $j=-\frac{1}{2}+i\rho, \rho \in \mathbb{R}$ for $k>2$ corresponds to 
the eigenvalue of the operator
$$\mathsf{j}=-\frac{1}{2}+ib^{-1}\mathsf{p}_{1}$$ and may
be interpreted as the spin of the $SL(2,\mathbb{R})$ representation in the principal
continuous series. We may furthermore
observe that the operator $a_{0}=-ib\mathsf{j}=\mathsf{p}_{1}+\frac{ib}{2}$ corresponds to the momentum zero mode of the Euclidean free boson $\phi_{1}(z),\bar{\phi}_{1}(\bar{z})$, where the shift is due to the presence of the background charge.

The Fock module $\mathcal{F}^{j}_{mn}\otimes \bar{\mathcal{F}}^{j}_{mn}$
 is spanned
by acting with the $\mathsf{a}^{(k)}_{-n},\mathsf{b}^{(k)}_{-n}$
on the vacuum state $|jmn\rangle$, which is annihilated
by the oscillators $\mathsf{a}^{(k)}_{n},\mathsf{b}^{(k)}_{n}$ with
 $n>0$.
For Fock states $\mathsf{f}$ in the Fock module $\mathcal{F}^{j}_{mn}\otimes
\bar{\mathcal{F}}^{j}_{mn}$ we introduce the notation $|jmn,\mathsf{f}\rangle$,
where we identify $|jmn\rangle\equiv |jmn,\Omega\rangle$, with $\Omega$ being
the Fock vacuum, and $\mathsf{a}^{(k)}_{-n}|jmn\rangle=|jmn,\mathsf{a}^{(k)}_{-n}\Omega\rangle$. 

We conclude this subsection by considering the spectrum
of the $\mathsf{L}_{0},\bar{\mathsf{L}}_{0}$ generators which, taking into
account the free field representation of the energy momentum tensor,
may be written
\be
\mathsf{L}_{0}=\mathsf{p}_{1}^{2}+\frac{b^2}{4}+(\mathsf{p}_{2}^{L})^{2}+2\sum_{k>0} \big (
\mathsf{a}_{-k}^{(1)}\mathsf{a}_{k}^{(1)}+\mathsf{a}_{-k}^{(2)}\mathsf{a}_{k}^{(2)}
\big )\\
\bar{\mathsf{L}_{0}}=\mathsf{p}_{1}^{2}+\frac{b^2}{4}+(\mathsf{p}_{2}^{R})^{2}+2\sum_{k>0} \big (
\mathsf{b}_{-k}^{(1)}\mathsf{b}_{k}^{(1)}+\mathsf{b}_{-k}^{(2)}\mathsf{b}_{k}^{(2)}
\big )
\ee
It immediately follows that their spectra are given by
\be
\mathrm{spec}\, \mathsf{L}_{0}&=&\bigg \{-\frac{j(j+1)}{k-2}+\frac{(m+nk)^{2}}{4k}+N \bigg |\,\, j\in -\frac{1}{2}+i\mathbb{R};m,n \in \mathbb{Z}, N \in \mathbb{N}\cup \{0\} \bigg \}\quad\\
\mathrm{spec}\, \bar{\mathsf{L}}_{0}&=&\bigg \{-\frac{j(j+1)}{k-2}+\frac{(m-nk)^{2}}{4k}+\bar{N}\bigg |\,\, j\in -\frac{1}{2}+i\mathbb{R};m,n \in \mathbb{Z}, \bar{N} \in \mathbb{N}\cup \{0\} \bigg \}\quad
\ee
This coincides with the spectra of the $\mathsf{L}_0,\bar{\mathsf{L}}_0$ generators derived in
\cite{dijkgraafverlinde} in the point-particle limit.

\subsection*{Quantum counterparts of Euclidean black hole fields}

We now wish to construct the quantum counterparts of the classical
building blocks $f_{1}(x_{+}),\\f_{2}(x_{+})$.
The basic building blocks $\mathsf{E}_{\alpha}^{(k)}(x_{+})=:e^{2\alpha\phi_{k}(x_{+})}:$ of this construction 
are the normal ordered exponentials which are discussed
 in appendix \ref{A}. As the quantum counterpart of $f_{1}$ we define
\be
\mathsf{f}_{1}(x_{+})=\mathsf{E}^{(1)}_{\frac{b}{2}}(x_{+})\mathsf{E}^{(2)}_{\frac{ib\eta}{2}}(x_{+})
\ee
Up to the chosen deformation of the coupling of $\phi_{1}$ in the
exponential, the
relation to the classical building block is quite obvious.
The construction of the quantum counterpart of building block $f_{2}(x_{+})$
 is a little bit
more involved. Let us first define the quantum counterpart
of the classical screening charge $A(x_{+})$.
In analogy to the classical theory we require the quantized screening
charge $\mathsf{A}(x_{+})$ to satisfy the differential equation
\be\label{screendiff}
\partial \mathsf{A}(x_{+})=\mathsf{V}(x_{+})
\ee
where $\mathsf{V}(x_{+})$  is the screening current which
in terms of chiral free fields is given by
\be
\mathsf{V}(x_{+})=\frac{1}{\sqrt{k}}:(\eta \partial \phi_{1}(x_{+})+i\partial \phi_{2}(x_{+}))e^{-2b\phi_{1}(x_{+})}:
\ee
The deformation of the coupling in the exponential as compared to the classical expression ($\frac{1}{\sqrt{k}}\rightarrow \frac{1}{\sqrt{k-2}}$)
is required to ensure vanishing of the conformal dimension of $\mathsf{A}(x_{+})$. The deformation parameter $\eta$  ensures that
the screening current $\mathsf{V}$ obeys a simple (parafermionic)
 exchange relation
with itself.

The solution to the differential equation (\ref{screendiff}) is unique if we
require monodromy to be preserved under integration. Taking into
account the odering of the zero-modes, we find the monodromy of $\mathsf{V}(x_{+})$
to be given by
\be\label{monov}
\mathsf{V}(x_{+}+2\pi)=e^{-4\pi b(\mathsf{p}_{1}-\frac{ib}{2})}\mathsf{V}
(x_{+})
\ee
The unique solution to (\ref{screendiff}) preserving monodromy is then
given by
\be
\mathsf{A}(x_{+})=\frac{e^{2\pi b (
\mathsf{p}_{1}-\frac{ib}{2})}}{2i\sin \pi b (2i\mathsf{p}_{1}+b)}
\mathsf{Q}(x_{+})
\ee
where
\be
\mathsf{Q}(x_{+})=\int\limits_{0}^{2\pi}d \varphi\,\mathsf{V}(x_{+}+\varphi)
\ee
The quantum counterpart of $f_{2}$ may then be defined as
\be
\mathsf{f}_{2}(x_{+})=\mathsf{A}(x_{+})\mathsf{f}_{1}(x_{+})
\ee
However, we have to analyze carefully the arising short-distance singularities
upon forming this product. To this purpose introduce the bilocal field
\be
\mathsf{f}_{2}(x_{+}',x_{+})=\mathsf{A}(x_{+}')\mathsf{f}_{2}(x_{+})
\ee
and analyze its behaviour as $x_{+}'$ approaches $x_{+}$.
We first integrate the total derivative appearing in the integrand of
$\mathsf{A}(x_{+})$ and obtain
\be
\mathsf{f}_{2}(x_{+}',x_{+})=-\frac{\eta^{2}}{2}\mathsf{E}^{(1)}_{-b}(x_{+}')
\mathsf{f}_{1}(x_{+})+ \frac{e^{2\pi b (
\mathsf{p}_{1}-\frac{ib}{2})}}{2i\sin \pi b (2i\mathsf{p}_{1}+b)}
\tilde{\mathsf{Q}}(x_{+})\mathsf{f}_{1}(x_{+})
\ee
with
\be
\tilde{\mathsf{Q}}(x_{+})=\frac{i}{\sqrt{k}}\int \limits_{0}^{2\pi}d\varphi\,:\partial\phi_{2}(x_{+}+\varphi)e^{-2b\phi_{1}(x_{+}+\varphi)}:
\ee
The short-distance behaviour of the first summand on the r.h.s.
is of special interest. According to eq. (\ref{shortdis}) it is given
by
\be\label{blabba}
\mathsf{E}^{(1)}_{-b}(x_{+}')
\mathsf{f}_{1}(x_{+})=e^{\frac{i\pi b^{2}}{2}\epsilon(x_{+}'-x_{+})}
|1-e^{-i(x_{+}'-x_{+})}|^{b^{2}}:\mathsf{E}^{(1)}_{-b}(x_{+}')
\mathsf{f}_{1}(x_{+}):
\ee
Here $\epsilon(x)$ denotes the stairstep function defined by
\be
\epsilon(x)=2n+1,\,\,2\pi n<x<2\pi(n+1),\,\,n\in \mathbb{Z}
\ee
For $k>2$ resp. real $b$ from \ref{blabba} we read off
\be
\lim_{x_{+}'\rightarrow x_{+}} \mathsf{E}^{(1)}_{-b}(x_{+}')
\mathsf{f}_{1}(x_{+})=0
\ee
which leads us to conclude
\be
\mathsf{f}_{2}(x_{+})=\lim_{x_{+}'\rightarrow x_{+}} \mathsf{A}(x_{+}')
\mathsf{f}_{2}(x_{+})=\frac{e^{2\pi b (
\mathsf{p}_{1}-\frac{ib}{2})}}{2i\sin \pi b (2i\mathsf{p}_{1}+b)}
\tilde{\mathsf{Q}}(x_{+}')\mathsf{f}_{1}(x_{+})
\ee
This may be seen as a kind of ``gauge degree of
freedom'' in the sense that adding a total derivative of
the dimensionless field $e^{-2b\phi_{1}}$ to the
integrand of the screening charge $\tilde{\mathsf{Q}}$
leaves physics unaltered as $\mathsf{f}_{2}$ remains invariant
under the ``gauge transformation'', i.e. upon constructing
the building block $\mathsf{f}_{2}$, the screening
charges $\mathsf{Q},\tilde{\mathsf{Q}}$ are equivalent.
 We will benefit
from this kind of ``gauge invariance'' when we calculate
the exchange algebra of the building blocks in the next subsection.
\\\\
Note that our definition of $\mathsf{f}_{2}$ makes sense
if the integrand of the screening charge $\mathsf{Q}$ develops
an integrable short-distance singularity with $\mathsf{f}_{1}$.
Our result is that $\mathsf{f}_{2}$ is well-defined for $k>2$, in which case
 the following identity holds:
\be\label{Qf}
\mathsf{Q}(x_{+})\mathsf{f}_{1}(x_{+})=e^{\frac{i\pi b^{2}}{2}}\int \limits_{0}^{2\pi}
d\varphi |1-e^{-i\varphi}|^{b^{2}}\, :\mathsf{V}(x_{+}+\varphi)\mathsf{f}_{1}(x_{+}):
\ee
It remains to define the quantum counterparts of the antichiral
building blocks. Quite obviously, the quantum counterpart of
$\bar{f}_{1}(x_{+})$ should be defined as
\be
\bar{\mathsf{f}}_{1}(x_{-})=\bar{\mathsf{E}}{(1)}_{\frac{b}{2}}(x_{-})\bar{\mathsf{E}}^{(2)}_{\frac{ib\eta}{2}}(x_{-})
\ee
The quantum counterpart of the antichiral building block $\bar{f}_{2}(x_{-})$
will then be defined as 
\be
\bar{\mathsf{f}}_{2}(x_{-})=\bar{\mathsf{f}}_{1}(x_{-})\bar{\mathsf{A}}(x_{-})
\ee
where the antichiral screening charge satisfies the diffential equation
\be
\partial\bar{\mathsf{A}}(x_{-})=\bar{\mathsf{V}}(x_{-})=
\frac{1}{\sqrt{k}}:(\eta \partial \bar{\phi}_{1}(x_{-})+i\partial \bar{\phi}_{2}(x_{-}))e^{-2b\bar{\phi}_{1}(x_{-})}:
\ee
The monodromy preserving solution to the above equation is given by
\be
\bar{\mathsf{A}}(x_{-})=\int \limits_{0}^{2\pi}
d\varphi\, \bar{\mathsf{V}}(x_{-}+\varphi)
\frac{e^{2\pi b (\mathsf{p}_{1}+\frac{ib}{2})}}{2i\sin \pi b (2i\mathsf{p}_{1}-b)}
\ee
Again, the so defined building block $\bar{\mathsf{f}}_{2}$ will be well defined
for $k>2$, in case of which we have
\be
\bar{\mathsf{f}}_{1}(x_{-})\int \limits_{0}^{2\pi}
d\varphi\, \bar{\mathsf{V}}(x_{-}+\varphi)=e^{-\frac{i\pi b^{2}}{2}}
\int \limits_{0}^{2\pi}d\varphi\, |1-e^{i\varphi}|^{b^{2}}
:\bar{\mathsf{f}}_{1}(x_{-})\bar{\mathsf{V}}(x_{-}+\varphi):
\ee
Note that the respective operator orderings chosen for 
the building blocks $\mathsf{f}_{2},\bar{\mathsf{f}}_{2}$
are fixed by requiring locality of the quantum counterpart of the
fundamental field $u(\sigma,t)$, which,  having identified the
 quantum counterparts of our building blocks, may be written
\be
\mathsf{u}(\sigma,t)=\mathsf{f}_{1}(x_{+})e^{-b\mathsf{q}_{1}}\bar{\mathsf{f}}_{1}(x_{-})
-\mathsf{f}_{2}(x_{+})e^{b\mathsf{q}_{1}}\bar{\mathsf{f}}_{2}(x_{-})
\ee
The quantum counterpart of the fundamental field $\bar{u}(\sigma,t)$
is then defined by $\bar{\mathsf{u}}(\sigma,t)=\mathsf{u}^{\dagger}(\sigma,t)$.

\subsection*{Conformal covariance}
The fundamental quantum fields $\mathsf{u},\bar{\mathsf{u}}$
will transform covariantly, as primary fields of
the Virasoro algebra, given that the (anti)chiral
building blocks $\mathsf{f}_{k},\bar{\mathsf{f}}_{k}$
transform covariantly with equal conformal dimensions.
This in turn holds if the screening charges $\mathsf{Q}(x_{+}),
\bar{\mathsf{Q}}(x_{-})$ have vanishing conformal dimensions,
which we will verify now. First of all
one may note that the building block $\mathsf{f}_{1}$
transforms due to (\ref{trans}) as
\be
[\mathsf{L}_{n},\mathsf{f}_{1}(x_{+})]=
e^{inx_{+}}(-i\partial_{+}+n\Delta)\mathsf{f}_{1}(x_{+})
\ee
with conformal dimension
\be\label{confdim}
\Delta=-\frac{3/4}{k-2}+\frac{1}{4k}
\ee
The building block $\mathsf{f}_{2}(x_{+})$ will then transform
covariantly with the same conformal dimension given that
the screening charge $\mathsf{Q}(x_{+})$ has vanishing 
conformal dimension. Indeed, the screening current $\mathsf{V}(x_{+})$
is primary of conformal dimension equal to one, i.e.
it obeys the commutation relation
\be
[\mathsf{L}_{n},\mathsf{V}(x_{+})]=e^{inx_{+}}(-i\partial_{+}+n)\mathsf{V}(x_{+})
\ee
which implies that the screening charge $\mathsf{Q}$ transforms
as a primary field of vanishing dimension according to
\be
[\mathsf{L}_{n},\mathsf{Q}(x_{+})]=
-ie^{inx_{+}}\partial_{+}\mathsf{Q}(x_{+})
\ee
One may argue analogously for the antichiral building blocks.
It immediately follows that $\mathsf{u}$ is primary,
\be
[\mathsf{L}_{n},\mathsf{u}(\sigma,t)]=e^{inx_{+}}(-i\partial_{+}+n\Delta)\mathsf{u}(\sigma,t)
\ee
\be
[\bar{\mathsf{L}}_{n},\mathsf{u}(\sigma,t)]=e^{inx_{-}}(-i\partial_{-}+n\Delta)\mathsf{u}(\sigma,t)
\ee
w.r.t. the chiral and antichiral copies of the Virasoro algebra,
with conformal dimension $\Delta$ given by (\ref{confdim}).

\subsection*{The braid algebra and proof of locality}

Locality of the quantum field $\mathsf{u}$ will be established
 by means of an exchange algebra satisfied by the chiral building
 blocks $\mathsf{f}_{1},\mathsf{f}_{2}$ and their antichiral companions.
We claim that the following relations hold ($i \not= j$):
\be\label{exchi}
\mathsf{f}_{i}(\sigma_{1})\mathsf{f}_{j}(\sigma_{2})=
\mathsf{f}_{j}(\sigma_{2})\mathsf{f}_{i}(\sigma_{1})C^{ij}_{ji}(\sigma_{2}-\sigma_{1},\mathsf{j})+
\mathsf{f}_{i}(\sigma_{2})\mathsf{f}_{j}(\sigma_{1})C^{ij}_{ij}(\sigma_{2}-\sigma_{1},\mathsf{j})
\ee
and correspondingly for the antichiral building blocks
\be\label{exanti}
\bar{\mathsf{f}}_{i}(-\sigma_{1})\bar{\mathsf{f}}_{j}(-\sigma_{2})=
\bar{\mathsf{f}}_{j}(-\sigma_{2})\bar{\mathsf{f}}_{i}(-\sigma_{1})\bar{C}^{ij}_{ji}(\sigma_{1}-\sigma_{2},\mathsf{j})+
\bar{\mathsf{f}}_{i}(-\sigma_{2})\bar{\mathsf{f}}_{j}(-\sigma_{1})\bar{C}^{ij}_{ij}(\sigma_{1}-\sigma_{2},\mathsf{j})
\ee
The coefficients $C^{ij}_{kl}$ of the braiding matrix are given by
\begin{align}\label{braidcoeff1}
C_{21}^{12}(\sigma,\mathsf{j})&=q^{(\frac{1}{k}-1)\epsilon(\sigma)}\,\,\frac{[2\mathsf{j}+2]_{q}[2\mathsf{j}]_{q}}{[2\mathsf{j}+1]_{q}^{2}}
  &  C_{12}^{21}(\sigma,\mathsf{j})&=q^{(\frac{1}{k}-1)\epsilon(\sigma)}  \nonumber \\
C_{12}^{12}(\sigma,\mathsf{j})&=q^{\frac{1}{k}\epsilon(\sigma)}\,\,\frac{q^{\,2\mathsf{j}\epsilon(\sigma)}}{[2\mathsf{j}+1]_{q}}  &  C_{21}^{21}(\sigma,\mathsf{j})&=-q^{\frac{1}{k}\epsilon(\sigma)}\,\,\frac{q^{(-2\mathsf{j}-2)\epsilon(\sigma)}}{[2\mathsf{j}+1]_{q}}
\end{align}
Here we have introduced q-deformed
numbers defined by ($q=e^{i\pi b^{2}}$)
\be
[x]_{q}=\frac{q^{x}-q^{-x}}{q-q^{-1}}
\ee
The coefficients of the antichiral exchange relations are related to
the coeffiecients of the corresponding chiral exchange relations in a simple
way:
\be\label{relchiantichi}
\bar{C}^{12}_{12}=C^{12}_{12}, \quad \bar{C}^{21}_{21}=C^{21}_{21}, \quad
\bar{C}^{12}_{21}=C^{21}_{12}, \quad \bar{C}^{21}_{12}=C^{12}_{21}
\ee
The exchange relations simplify for $i=j$ where we have
\be\label{ff}
\mathsf{f}_{i}(\sigma_{1})\mathsf{f}_{i}(\sigma_{2})&=&q^{\frac{1}{k}\epsilon(\sigma_{2}-\sigma_{1})}\mathsf{f}_{i}(\sigma_{2})\mathsf{f}_{i}(\sigma_{1})\\
\bar{\mathsf{f}}_{i}(-\sigma_{1})\bar{\mathsf{f}}_{i}(-\sigma_{2})&=&q^{\frac{1}{k}\epsilon(\sigma_{1}-\sigma_{2})}\bar{\mathsf{f}}_{i}(-\sigma_{2})\bar{\mathsf{f}}_{i}(-\sigma_{1})
\ee
The details of the proof of the braid algebra can be found in appendix C.
Having established the braid algebra, we may now  prove locality
of the Euclidean black hole fields. First of all note
that application of the braid algebra on the r.h.s. of (\ref{exchi})
leads to the following consistency conditions on the coefficients
of the braid relations:
\be\label{consistent}
C^{ji}_{ij}(\sigma,\mathsf{j})C^{ij}_{ji}(-\sigma,\mathsf{j})+C^{ij}_{ij}(\sigma,\mathsf{j})C^{ij}_{ij}(-\sigma,\mathsf{j})=1 \nonumber \\
C^{ji}_{ji}(\sigma,\mathsf{j})C^{ij}_{ji}(-\sigma,\mathsf{j})+C^{ij}_{ji}(\sigma,\mathsf{j})C^{ij}_{ij}(-\sigma,\mathsf{j})=0
\ee
Corresponding relations hold for the coefficients $\bar{C}^{ij}_{kl}$
of the antichiral braid algebra. We will now
see that the consistency conditions (\ref{consistent}) and the relation (\ref
{relchiantichi}) between
the coefficients of the chiral and antichiral braid algebra 
suffice to establish locality of the quantum fields $u,\bar{u}$. 
Consider the operator product
\be\label{uop}
\mathsf{u}(\sigma_{1},0)\mathsf{u}(\sigma_{2},0)=\mathsf{f}_{11}(\sigma_{1},\sigma_{2})-\mathsf{f}_{12}(\sigma_{1},\sigma_{2})-\mathsf{f}_{21}(\sigma_{1},\sigma_{2})+\mathsf{f}_{22}(\sigma_{1},\sigma_{2})
\ee
where we have introduced bilocal fields $\mathsf{f}_{ij}$ according to
\be
\mathsf{f}_{11}(\sigma_{1},\sigma_{2})&=&\mathsf{f}_{1}(\sigma_{1})\,e^{-b\mathsf{q}_{1}}\,\bar{\mathsf{f}}_{1}(-\sigma_{1})\,\,\mathsf{f}_{1}(\sigma_{2})\,e^{-b\mathsf{q}_{1}}\,\bar{\mathsf{f}}_{1}(-\sigma_{2}) \nonumber \\
\mathsf{f}_{22}(\sigma_{1},\sigma_{2})&=&\mathsf{f}_{2}(\sigma_{1})\,e^{b\mathsf{q}_{1}}\,\bar{\mathsf{f}}_{2}(-\sigma_{1})\,\,\mathsf{f}_{2}(\sigma_{2})\,e^{b\mathsf{q}_{1}}\,\bar{\mathsf{f}}_{2}(-\sigma_{2}) \nonumber \\
\mathsf{f}_{12}(\sigma_{1},\sigma_{2})&=&\mathsf{f}_{1}(\sigma_{1})\,e^{-b\mathsf{q}_{1}}\,\bar{\mathsf{f}}_{1}(-\sigma_{1})\,\,\mathsf{f}_{2}(\sigma_{2})\,e^{b\mathsf{q}_{1}}\,\bar{\mathsf{f}}_{2}(-\sigma_{2}) \nonumber \\
\mathsf{f}_{21}(\sigma_{1},\sigma_{2})&=&\mathsf{f}_{2}(\sigma_{1})\,e^{b\mathsf{q}_{1}}\,\bar{\mathsf{f}}_{2}(-\sigma_{1})\,\,\mathsf{f}_{1}(\sigma_{2})\,e^{-b\mathsf{q}_{1}}\,\bar{\mathsf{f}}_{1}(-\sigma_{2})
\ee
and we wish to show $\mathsf{u}(\sigma_{1},0)\mathsf{u}(\sigma_{2},0)=\mathsf{u}(\sigma_{2},0)\mathsf{u}(\sigma_{1},0)$. First of
all let us note that by means of (\ref{ff}) it immediately follows that
\be\label{fiifii}
\mathsf{f}_{11}(\sigma_{1},\sigma_{2})&=&\mathsf{f}_{11}(\sigma_{2},\sigma_{1}) \nonumber \\
\mathsf{f}_{22}(\sigma_{1},\sigma_{2})&=&\mathsf{f}_{22}(\sigma_{2},\sigma_{1})
\ee
Exchange relations for the remaining terms in (\ref{uop}) can be obtained by means of the exchange algebra (\ref{exchi}) resp. (\ref{exanti}), but it is first of all useful
to note that the operator combinations $e^{-b\mathsf{q}_{1}}\,\bar{\mathsf{f}}_{1}(-\sigma_{1})$, $\mathsf{f}_{2}(\sigma_{2})\,e^{b\mathsf{q}_{1}}$, $e^{b\mathsf{q}_{1}}\,\bar{\mathsf{f}}_{2}(-\sigma_{1})$ and $\mathsf{f}_{1}(\sigma_{2})e^{-b\mathsf{q}_{1}}$ appearing in $\mathsf{f}_{12}, \mathsf{f}_{21}$ have no dependence on the zero-mode $\mathsf{q}_{1}$. Therefore, the chiral objects
commute with the corresponding antichiral objects. We may therefore
change their ordering and rewrite the bilocal fields $\mathsf{f}_{12}, \mathsf{f}_{21}$ according to
\be
\mathsf{f}_{12}(\sigma_{1},\sigma_{2})&=&\mathsf{f}_{1}(\sigma_{1})\mathsf{f}_{2}(\sigma_{2})\times\bar{\mathsf{f}}_{1}(-\sigma_{1})\bar{\mathsf{f}}_{2}(-\sigma_{2}) \nonumber \\
\mathsf{f}_{21}(\sigma_{1},\sigma_{2})&=&\mathsf{f}_{2}(\sigma_{1})\mathsf{f}_{1}(\sigma_{2})\times\bar{\mathsf{f}}_{2}(-\sigma_{1})\bar{\mathsf{f}}_{1}(-\sigma_{2})
\ee
We then apply the exchange algebra (\ref{exchi}) and (\ref{exanti}) on the r.h.s. and find
\begin{multline}
\mathsf{f}_{12}(\sigma_{1},\sigma_{2})=\\ \bigg (\mathsf{f}_{2}(\sigma_{2})\mathsf{f}_{1}(\sigma_{1})\,C_{21}^{12}(\sigma_{2}-\sigma_{1}\mathsf{j})+\mathsf{f}_{1}(\sigma_{2})\mathsf{f}_{2}(\sigma_{1})\,C_{12}^{12}(\sigma_{2}-\sigma_{1},\mathsf{j})\bigg ) \\  \times
\bigg (\bar{\mathsf{f}}_{2}(-\sigma_{2})\bar{\mathsf{f}}_{1}(-\sigma_{1})\,\bar{C}_{21}^{12}(\sigma_{1}-\sigma_{2},\mathsf{j})+\bar{\mathsf{f}}_{1}(-\sigma_{2})\bar{\mathsf{f}}_{2}(-\sigma_{1})\,\bar{C}_{12}^{12}(\sigma_{1}-\sigma_{2},\mathsf{j})\bigg )
\end{multline}
Now, since the $C^{ij}_{kl}(\sigma,\mathsf{j})$, $\bar{C}^{ij}_{kl}(\sigma,\mathsf{j})$ commute with the operator products $\mathsf{f}_{1}\mathsf{f}_{2}$, $\mathsf{f}_{2}\mathsf{f}_{1}$ and $\bar{\mathsf{f}}_{1}\bar{\mathsf{f}}_{2}$ ,  $\bar{\mathsf{f}}_{2}\bar{\mathsf{f}}_{1}$ this may be rewritten as
\begin{multline}\label{f12}
\mathsf{f}_{12}(\sigma_{1},\sigma_{2})=\\ \mathsf{f}_{21}(\sigma_{2},\sigma_{1})\,\,C_{21}^{12}(\sigma_{2}-\sigma_{1}\mathsf{j})\bar{C}_{21}^{12}(\sigma_{1}-\sigma_{2},\mathsf{j})+\mathsf{f}_{12}(\sigma_{2},\sigma_{1})\,\,C_{12}^{12}(\sigma_{2}-\sigma_{1},\mathsf{j})\bar{C}_{12}^{12}(\sigma_{1}-\sigma_{2},\mathsf{j})\\
+\mathsf{f}_{1}(\sigma_{2})\mathsf{f}_{2}(\sigma_{1})\bar{\mathsf{f}}_{2}(-\sigma_{2})\bar{\mathsf{f}}_{1}(-\sigma_{1})\,\,C_{12}^{12}(\sigma_{2}-\sigma_{1},\mathsf{j})\bar{C}_{21}^{12}(\sigma_{1}-\sigma_{2},\mathsf{j})\\
+\mathsf{f}_{2}(\sigma_{2})\mathsf{f}_{1}(\sigma_{1})\bar{\mathsf{f}}_{1}(-\sigma_{2})\bar{\mathsf{f}}_{2}(-\sigma_{1})\,\,C_{21}^{12}(\sigma_{2}-\sigma_{1}\mathsf{j})\bar{C}_{12}^{12}(\sigma_{1}-\sigma_{2},\mathsf{j})
\end{multline}
Correspondingly, we find the exchange relation
\begin{multline}\label{f21}
\mathsf{f}_{21}(\sigma_{1},\sigma_{2})=\\ \mathsf{f}_{12}(\sigma_{2},\sigma_{1})\,\,C_{12}^{21}(\sigma_{2}-\sigma_{1}\mathsf{j})\bar{C}_{12}^{21}(\sigma_{1}-\sigma_{2},\mathsf{j})+\mathsf{f}_{21}(\sigma_{2},\sigma_{1})\,\,C_{21}^{21}(\sigma_{2}-\sigma_{1},\mathsf{j})\bar{C}_{21}^{21}(\sigma_{1}-\sigma_{2},\mathsf{j})\\
+\mathsf{f}_{2}(\sigma_{2})\mathsf{f}_{1}(\sigma_{1})\bar{\mathsf{f}}_{1}(-\sigma_{2})\bar{\mathsf{f}}_{2}(-\sigma_{1})\,\,C_{21}^{21}(\sigma_{2}-\sigma_{1},\mathsf{j})\bar{C}_{12}^{21}(\sigma_{1}-\sigma_{2},\mathsf{j})\\
+\mathsf{f}_{1}(\sigma_{2})\mathsf{f}_{2}(\sigma_{1})\bar{\mathsf{f}}_{2}(-\sigma_{2})\bar{\mathsf{f}}_{1}(-\sigma_{1})\,\,C_{12}^{21}(\sigma_{2}-\sigma_{1}\mathsf{j})\bar{C}_{21}^{21}(\sigma_{1}-\sigma_{2},\mathsf{j})
\end{multline}
It follows that the operator product $\mathsf{u}(\sigma_{1},0)\mathsf{u}(\sigma_{2},0)$ may be rewritten according to
\begin{multline}
\mathsf{u}(\sigma_{1},0)\mathsf{u}(\sigma_{2},0)=\mathsf{f}_{11}(\sigma_{2},\sigma_{1})+\mathsf{f}_{22}(\sigma_{2},\sigma_{1}) \\
-\mathsf{f}_{12}(\sigma_{2},\sigma_{1})\bigg(C_{12}^{12}(\sigma_{2}-\sigma_{1},\mathsf{j})\bar{C}_{12}^{12}(\sigma_{1}-\sigma_{2},\mathsf{j})+C_{12}^{21}(\sigma_{2}-\sigma_{1}\mathsf{j})\bar{C}_{12}^{21}(\sigma_{1}-\sigma_{2},\mathsf{j}) \bigg )\\
-\mathsf{f}_{21}(\sigma_{2},\sigma_{1})\bigg (C_{21}^{12}(\sigma_{2}-\sigma_{1}\mathsf{j})\bar{C}_{21}^{12}(\sigma_{1}-\sigma_{2},\mathsf{j})+C_{21}^{21}(\sigma_{2}-\sigma_{1},\mathsf{j})\bar{C}_{21}^{21}(\sigma_{1}-\sigma_{2},\mathsf{j}) \bigg) \\
-\mathsf{f}_{1}(\sigma_{2})\mathsf{f}_{2}(\sigma_{1})\bar{\mathsf{f}}_{2}(-\sigma_{2})\bar{\mathsf{f}}_{1}(-\sigma_{1}) \bigg (C_{12}^{12}(\sigma_{2}-\sigma_{1},\mathsf{j})\bar{C}_{21}^{12}(\sigma_{1}-\sigma_{2},\mathsf{j})+C_{12}^{21}(\sigma_{2}-\sigma_{1}\mathsf{j})\bar{C}_{21}^{21}(\sigma_{1}-\sigma_{2},\mathsf{j}) \bigg )\\
-\mathsf{f}_{2}(\sigma_{2})\mathsf{f}_{1}(\sigma_{1})\bar{\mathsf{f}}_{1}(-\sigma_{2})\bar{\mathsf{f}}_{2}(-\sigma_{1})\bigg (C_{21}^{12}(\sigma_{2}-\sigma_{1}\mathsf{j})\bar{C}_{12}^{12}(\sigma_{1}-\sigma_{2},\mathsf{j})+C_{21}^{21}(\sigma_{2}-\sigma_{1},\mathsf{j})\bar{C}_{12}^{21}(\sigma_{1}-\sigma_{2},\mathsf{j})\bigg )\\
\end{multline}
The field $\mathsf{u}$ thus satisfies the canonical equal-time commutation relation given that the following equations are satisfied :
\be\label{comcon}
C_{12}^{12}(\sigma_{2}-\sigma_{1},\mathsf{j})\bar{C}_{12}^{12}(\sigma_{1}-\sigma_{2},\mathsf{j})+C_{12}^{21}(\sigma_{2}-\sigma_{1},\mathsf{j})\bar{C}_{12}^{21}(\sigma_{1}-\sigma_{2},\mathsf{j})&=& 1 \nonumber \\
C_{21}^{12}(\sigma_{2}-\sigma_{1},\mathsf{j})\bar{C}_{21}^{12}(\sigma_{1}-\sigma_{2},\mathsf{j})+C_{21}^{21}(\sigma_{2}-\sigma_{1},\mathsf{j})\bar{C}_{21}^{21}(\sigma_{1}-\sigma_{2},\mathsf{j})&=& 1 \nonumber \\
C_{12}^{12}(\sigma_{2}-\sigma_{1},\mathsf{j})\bar{C}_{21}^{12}(\sigma_{1}-\sigma_{2},\mathsf{j})+C_{12}^{21}(\sigma_{2}-\sigma_{1},\mathsf{j})\bar{C}_{21}^{21}(\sigma_{1}-\sigma_{2},\mathsf{j})&=& 0 \nonumber \\
C_{21}^{12}(\sigma_{2}-\sigma_{1},\mathsf{j})\bar{C}_{12}^{12}(\sigma_{1}-\sigma_{2},\mathsf{j})+C_{21}^{21}(\sigma_{2}-\sigma_{1},\mathsf{j})\bar{C}_{12}^{21}(\sigma_{1}-\sigma_{2},\mathsf{j})&=& 0
\ee
Taking into account the identities (\ref{relchiantichi}) relating the exchange algebra coefficients of the anitchiral building-blocks to those of the chiral building-blocks,
we find that the set of equations (\ref{comcon}) reduces to the consistency-conditions (\ref{consistent}), and the proof is complete.  
The proof of the equal time commutator $[\,\bar{\mathsf{u}}(\sigma_{1},t),\bar{\mathsf{u}}(\sigma_{2},t)\,]=0$ is then
trivial since $\bar{\mathsf{u}}(\sigma,t)=\mathsf{u}^{\dagger}(\sigma,t)$  .

\subsection*{Matrix elements}
In this subsection we will consider the matrix elements of the
Euclidean black hole fields between vacua $|jmn\rangle$ in 
the Fock module $\mathcal{F}^{j}_{mn}\otimes \mathcal{F}^{j}_{mn}$. Knowledge
of these vacuum matrix elements will be crucial for the
derivation of the reflection operator of the Euclidean black hole model to be
presented in the next subsection. Upon calculating the matrix
elements, we have to carefully keep track of the zero mode exponentials
which shift the values of $j,m,n$ when applied to the vacuum state $|jmn\rangle$.
Then, keeping in mind formula (\ref{Qf}) for $\mathsf{f}_{2}$ (and the corresponding
expression for $\bar{\mathsf{f}}_{2}$), it turns out
that the vacuum matrix element of $\mathsf{u}$ for $t=0$ may be written
\begin{multline}\label{umat}
\langle j'm'n'|\,\mathsf{u}(\sigma,0)\,|jmn\rangle =\\e^{in\sigma}\,\,\delta_{m',m+1}\,\delta_{nn'} \bigg (\delta(j'-j-\frac{1}{2})\,+\,\frac{1}{k^{2}}\,
\bigl (j-\frac{1}{2}(m+nk) \bigr ) \bigl (j-\frac{1}{2}(m-nk) \bigr )\,\,\mathcal{B}(j)\,\,\bar{\mathcal{B}}(j)\,\,\delta(j'-j+\frac{1}{2}) \bigg )
\end{multline}
where $\mathcal{B}(j),\bar{\mathcal{B}}(j)$ are given by
\be
\mathcal{B}(j)&=&\frac{e^{-i\pi b^{2}(2j+1)}}{2i\sin \pi b^{2}(2j+1)}\int \limits_{0}^{2\pi} d\varphi \,\,(1-e^{-i\varphi})^{b^{2}}\,e^{i\varphi\, b^{2}(2j+1)}\\
\bar{\mathcal{B}}(j)&=&\frac{e^{-2\pi i b^{2}j}}{2i\sin 2\pi b^{2}j}\int \limits_{0}^{2\pi} d\varphi \,\,(1-e^{i\varphi})^{b^{2}}\,e^{i\varphi\, b^{2}\,2j}
\ee
By a simple contour deformation, the integrals in the above equations
may be related to the integral representation of the Euler beta function,
giving the result
\be
\int \limits_{0}^{2\pi} d\varphi \,\,(1-e^{-i\varphi})^{b^{2}}\,e^{i\varphi\, b^{2}(2j+1)}&=&2\pi \Gamma(1+b^{2})\frac{e^{i \pi b^{2}(2j+1)}}
{\Gamma(1-2b^{2}j)\Gamma(1+b^{2}(2j+1))}\\
\int \limits_{0}^{2\pi} d\varphi \,\,(1-e^{i\varphi})^{b^{2}}\,e^{i\varphi\, b^{2}\,2j}&=&2\pi \Gamma(1+b^{2})\frac{e^{2i \pi b^{2}j}}
{\Gamma(1-2b^{2}j)\Gamma(1+b^{2}(2j+1))}
\ee
This finally yields for $\mathcal{B}(j),\bar{\mathcal{B}}(j)$ the
expressions
\be
\mathcal{B}(j)=i\,\Gamma(1+b^{2})\frac{\Gamma(-b^{2}(2j+1))}{\Gamma(1-2b^{2}j)}\,\,  , \qquad \bar{\mathcal{B}}(j)=-i\Gamma(1+b^{2})\frac{\Gamma(2b^{2}j)}{\Gamma(1+b^{2}(2j+1))}
\ee
where we have used the reflection property of the Gamma function, $\Gamma(x)
\Gamma(1-x)=\pi/\sin\pi x$.

\subsection*{The reflection operator}
The quantum counterpart of $S: \mathcal{P}^{F}_{\pm}\rightarrow \mathcal{P}^{F}_{\mp}$ should be a unitary operator that maps $\mathsf{p}_{1}$ into
$-\mathsf{p}_{1}$ (resp. $\mathsf{j}$ into $-\mathsf{j}-1$) but
leaves $\mathsf{u}$ invariant,
\be
\mathcal{S}\mathsf{u}\mathcal{S}^{-1}=\mathsf{u}
\ee
Consistency with the symmetry algebra of the model implies that
$\mathcal{S}$ commutes with the modes of the W-algebra currents, a suitable
ansatz for $\mathcal{S}$ is therefore
\be\label{ansatz}
\mathcal{S}=\mathcal{P}_{1}\mathsf{R}
\ee
where $\mathcal{P}_1$ denotes the parity operator on the zero mode
space of free boson $\Phi_{1}$, and $\mathsf{R}$ is an intertwiner
between representations of the W-algebra,
\be\label{intertwiner}
\mathsf{R}\,\mathsf{W}_{s,n}(\mathsf{j},\mathsf{m},\mathsf{n})=\mathsf{W}_{s,n}(-\mathsf{j}-1,\mathsf{m},\mathsf{n})\,
\mathsf{R}
\ee
The operator $\mathsf{R}$ will be unitary if the following
conditions are satisfied. First of all, since
$\mathsf{R}$ commutes with $\mathsf{L}_{0},\bar{\mathsf{L}}_{0}$,
the Fock vacuum $|jmn\rangle$ should be an eigenstate of $\mathsf{R}$,
i.e. we require
\be\label{eigenvalue}
\mathsf{R}|jmn\rangle=R(j,m,n)|jmn\rangle
\ee
The operator $\mathsf{R}$ will then be unitary iff $|R(j,m,n)|^{2}=1$
and the norm of
any state in the W-algebra module $\mathcal{W}^{j}_{mn}$ is invariant
under $j\rightarrow -j-1$, i.e. if we have
\begin{multline}\label{inva}
||\mathsf{W}_{s_1,-n_{1}}(j,m,n)\cdots \mathsf{W}_{s_k,-n_k}(j,m,n)\Omega||=\\
||\mathsf{W}_{s_1,-n_{1}}(-j-1,m,n)\cdots \mathsf{W}_{s_k,-n_k}(-j-1,m,n)\Omega||
\end{multline}
where the norm is induced by the scalar product in $\mathcal{W}^{j}_{mn}$ which is defined by means of the hermiticity
relations
\be
\mathsf{W}_{s,n}^{\dagger}=\mathsf{W}_{s,-n}
\ee
and the fact that the Fock vacua $|jmn\rangle$ are highest weight states
of the W-algebra, $\mathsf{W}_{s,n}|jmn\rangle=0,\,n>0$. In order
to prove (\ref{inva}), note that the statement is equivalent to invariance
of the eigenvalues of the zero-modes $\mathsf{W}_{s,0}$ under the
reflection $j\rightarrow -j-1$, which is proven in appendix A.  
Note that corresponding statements hold for the anitholomorphic
W-algebra module $\bar{\mathcal{W}}^{j}_{mn}$ spanned by acting with
 the $\bar{\mathsf{W}}_{s,-n}$ on the Fock vacuum $|jmn\rangle$.

Having established unitarity of $\mathsf{R}$, we wish to define
its action upon the Fock module $\mathcal{F}^{j}_{mn}$ spanned by 
acting with the oscillators $\mathsf{a}^{(k)}_{-n}$ on $|jmn\rangle$.
We will argue that the action of $\mathsf{R}$ is fixed up to
a phase if $\mathcal{W}^{j}_{mn}$ is isomorphic to $\mathcal{F}^{j}_{mn}$.
Clearly, any state in $\mathcal{W}^{j}_{mn}$ is also a Fock state
since the $\mathsf{W}_{s,n}$ may be expressed by the oscillators.
The reverse statement is certainly true if $\mathcal{W}^{j}_{mn}$
is irreducible, i.e. does not contain any null vectors
\footnote{It is possible to deduce the relevant result concerning the isomorphism
of $\mathcal{F}^j_{mn}$ and $\mathcal{W}^j_{mn}$ from the determinant
formula of \cite{dixon}.}.  We assume
that this holds at least for $j \in -\frac{1}{2}+i\mathbb{R}$.
Then, any Fock state $\mathsf{f}$ in $\mathcal{F}^{j}_{mn}$
 may be written
\be
\mathsf{f}=\mathcal{P}^{j}_{mn}(\mathsf{L}_{-n}(j,m,n),\mathsf{W}_{3,-n}(j,m,n))\Omega
\ee
where $\mathcal{P}^{j}_{mn}$ is some polynomial in the
$\mathsf{L}_{-n}(j,m,n),\mathsf{W}_{3,-n}(j,m,n)$ with coefficients depending
on the quantum numbers $j,m,n$.
Here we make the additional assumption that $\mathcal{W}^{j}_{mn}$
is already covered by acting with the $\mathsf{L}_{-n},\mathsf{W}_{3,-n}$
on the Fock vacuum.
Acting with the reflection operator $\mathsf{R}$ on the Fock state
$\mathsf{f}$, we find, using (\ref{intertwiner}) and (\ref{eigenvalue}),
\be
\mathsf{R}\mathsf{f}=R(j,m,n)\,\mathcal{P}^{j}_{mn}(\mathsf{L}_{-n}(-j-1,m,n),\mathsf{W}_{3,-n}(-j-1,m,n))\Omega
\ee
 Completely analogous, we may
define the action of $\mathsf{R}$ on Fock states in the Fock module 
 $\bar{\mathcal{F}}^{j}_{mn}$ spanned by acting with the
$\mathsf{b}^{(k)}_{-n}$ on $|jmn\rangle$.

 Therefore, due to symmetry considerations,
the operator $\mathsf{R}$ is uniquely characterized
by the reflection amplitude $R(j,m,n)$. In order to
determine $R(j,m,n)$ use the fact that $\mathcal{S}$ leaves
$\mathsf{u}$ invariant and consider the equation
\be
\langle j'm'n'|\mathcal{S}\mathsf{u}(\sigma,0)\mathcal{S}^{-1}|jmn\rangle
=\langle j'm'n'|\mathsf{u}(\sigma,0)|jmn\rangle
\ee
Using equations (\ref{ansatz}),(\ref{eigenvalue}) and the expression (\ref{umat}) for the
matrix elements of $\mathsf{u}$, we find by comparison
of the coefficients of the arising delta functions that
the reflection amplitude $R(j,m,n)$ solves the difference equations
\begin{multline}
R(-j-\frac{1}{2},m+1,n)R(j,m,n)=\\\frac{1}{k^{2}}\,\bigg (j-\frac{1}{2}(m+nk) \bigg ) \bigg (j-\frac{1}{2}(m-nk) \bigg )\,\Gamma^{2}(1+b^{2})
\frac{\Gamma(-b^{2}(2j+1))\Gamma(2b^{2}j)}{\Gamma(1-2b^{2}j)\Gamma(1+b^{2}(2j+1))}
\end{multline}
\begin{multline}
R(-j-\frac{3}{2},m+1,n)R(j,m,n)=k^{2} \bigg (-j-1-\frac{1}{2}(m+nk) \bigg )^{-1} \bigg (-j-1-\frac{1}{2}(m-nk) \bigg )^{-1}\\ \times \frac{1}{\Gamma^{2}(1+b^{2})}\frac{\Gamma(1+2b^{2}(j+1))\Gamma(1-b^{2}(2j+1))}{\Gamma(b^{2}(2j+1))\Gamma(-2b^{2}(j+1)}
\end{multline}
It is easy to see that these equations are solved by
\begin{multline}
R(j,m,n)=\\ \bigg ( \nu(k) \bigg )^{2j+1} \frac{\Gamma(j+1+\frac{1}{2}(m+nk))}{\Gamma(-j+\frac{1}{2}(m+nk))}\frac{\Gamma(j+1+\frac{1}{2}(m-nk))}{\Gamma(-j+\frac{1}{2}(m-nk))}\frac{\Gamma(-2j-1)}{\Gamma(2j+1)}\frac{\Gamma(1-\frac{2j+1}{k-2})}{\Gamma(1+\frac{2j+1}{k-2})}
\end{multline}
with
\be
\nu(k)=\frac{1}{k^{2}}\Gamma^{2}(\frac{1}{k-2})
\ee
It remains to verify $|R(j,m,n)|^2=1$. Since we consider only representations
where $j\in -\frac{1}{2}+i\mathbb{R}$, it follows that $j^{\ast}=-j-1$.
From this it can easily be inferred that $R(j,m,n)$ is unitary.

To conclude this subsection, let us consider the semiclassical limit $k\rightarrow\infty$ of our scattering amplitude. In this limit we have
\be
R(j,m,n) \overset{k\rightarrow \infty}{\sim}\frac{\Gamma(j+1+\frac{1}{2}(m+nk))}{\Gamma(-j+\frac{1}{2}(m+nk))}\frac{\Gamma(j+1+\frac{1}{2}(m-nk))}{\Gamma(-j+\frac{1}{2}(m-nk))}\frac{\Gamma(-2j-1)}{\Gamma(2j+1)}
\ee
This precisely coincides with the semiclassical results of \cite{dijkgraafverlinde}.

\section*{Acknowledgements}
I thank Dr. J. Teschner for fruitful discussions and valuable contributions to the paper.
 I furthermore thank Prof. R. Schrader for encouragement.

\appendix
\section{Normal ordered exponentials}\label{A}

The main ingredience of our construction of
the quantum counterparts of the Euclidean black hole fields are
(anti)chiral normal ordered exponentials. See for example
\cite{nicolai} for a more detailed explanation of the techniques
used here. The normal ordered exponentials are defined as
\be\label{noex}
\mathsf{E}_{\alpha}^{(k)}(x_{+})=\sum_{n=0}^{\infty} \frac{(2\alpha)^{n}}{n!}:\phi_{k}(x_{+})^{n}:\,, \qquad \bar{\mathsf{E}}_{\alpha}^{(k)}(x_{-})=\sum_{n=0}^{\infty} \frac{(2\alpha)^{n}}{n!}:\bar{\phi}_{k}(x_{-})^{n}:
\ee
where we impose the usual Wick-ordering for the oscillators
and symmetric normal-ordering for the zero-modes, i.e.
\be
:\mathsf{q}^{n}f(\mathsf{p}):=\frac{d^{n}}{d \alpha^{n}}(e^{\frac{\alpha}{2}\mathsf{q}} f(\mathsf{p})e^{\frac{\alpha}{2}\mathsf{q}})|_{\alpha=0}
\ee
Note that the definition (\ref{noex}) is equivalent to
\be
\mathsf{E}_{\alpha}^{(1)}(x_{+})=e^{\alpha \mathsf{q}_{1}}
\exp\bigg\{2i\alpha\sum\limits_{n<0} \frac{\mathsf{a}_n^{(1)}}{n}e^{-inx_{+}}\bigg\}e^{2\alpha\mathsf{p}_{1}x_{+}}\exp\bigg\{2i\alpha\sum\limits_{n>0} \frac{\mathsf{a}_n^{(1)}}{n}e^{-inx_{+}}\bigg\}e^{\alpha \mathsf{q}_{1}}
\ee
and
\be
\mathsf{E}_{\alpha}^{(2)}(x_{+})=e^{\alpha \mathsf{q}_{2}^L}
\exp\bigg\{2i\alpha\sum\limits_{n<0} \frac{\mathsf{a}_n^{(2)}}{n}e^{-inx_{+}}\bigg\}e^{2\alpha\mathsf{p}_{2}^L x_{+}}\exp\bigg\{2i\alpha\sum\limits_{n>0} \frac{\mathsf{a}_n^{(2)}}{n}e^{-inx_{+}}\bigg\}e^{\alpha \mathsf{q}_{2}^L}
\ee
Corresponding expressions exist for the antichiral exponentials. From now on we will suppress the 
discussion of the antichiral exponentials since all results for the chiral exponentials can directly be
applied to the antichiral counterparts.
Using Wick's theorem, one may show that the normal ordered exponentials have short-distance
behaviour
\be\label{shortdis}
\mathsf{E}_{\alpha}^{(k)}(x_{+})\mathsf{E}_{\beta}^{(k)}(y_{+})=e^{-i\pi \alpha \beta \epsilon(x_{+}-y_{+})}|1-e^{-i(x_{+}-y_{+})}|^{-2\alpha\beta}:\mathsf{E}_{\alpha}^{(k)}(x_{+})\mathsf{E}_{\beta}^{(k)}(y_{+}):
\ee
Here, as starting point for the application of Wick's theorem, we use the
fact the the chiral quantum field $\phi_{k}$
satisfies 
\be
\phi_{k}(x_{+})\phi_{l}(y_{+})&=&:\phi_{k}(x_{+})\phi_{l}(y_{+}):
-\frac{i\pi}{2}\delta_{kl}\epsilon^{+}(x_{+}-y_{+})
\ee
where the distribution $\epsilon^{+}$ is given by
\be
\epsilon^{+}(x)=\frac{1}{2}\epsilon(x)-\frac{i}{\pi}\ln|1-e^{-ix}|=\frac{x}{2\pi}-\frac{i}{\pi}\ln (1-e^{-ix})
\ee
Here, $\ln$ denotes the principal value of the logarithm.
From the above equation one easily derives that the normal-ordered
exponentials
 obey the exchange-relations
\be\label{exe}
\mathsf{E}_{\alpha}^{(k)}(x_{+})\mathsf{E}_{\beta}^{(k)}(y_{+})=e^{-2\pi i \alpha \beta \epsilon(x_{+}-y_{+})}\mathsf{E}_{\beta}^{(k)}(y_{+})\mathsf{E}_{\alpha}^{(k)}(x_{+})
\ee
The normal ordered exponentials are primary fields of the
Virasoro algebra and satisfy commutation relations with the generators
$\mathsf{L}_{n}$ of the form
\be\label{trans}
[\mathsf{L}_{n},\mathsf{E}_{\alpha}^{(k)}(x_{+})] &=& e^{inx_{+}}(-i\partial_{+}+
n\Delta^{(k)}_{\alpha})\mathsf{E}_{\alpha}^{(k)}(x_{+})
\ee
The conformal weights are given by
\be
\Delta^{(k)}_{\alpha}=\alpha(Q\delta_{k1}-\alpha)
\ee
where $Q=-b$ is the background charge of the theory.\\
 Causal bosonic quantum fields are obtained by composing chiral and antichiral
exponentials according to
\be
:e^{2\alpha \Phi_{1}(\sigma,t)}:&=&\mathsf{E}^{(1)}_{\alpha}(x_{+})e^{-2\alpha \mathsf{q}_{1}} \bar{\mathsf{E}}^{(1)}_{\alpha}(x_{-})\\
:e^{2i\beta \Phi_{2}(\sigma,t)}:&=&\mathsf{E}^{(2)}_{i\beta}(x_{+})\bar{\mathsf{E}}^{(2)}_{i\beta}(x_{-})
\ee
where we impose the restriction $\beta \in \mathbb{R}$ since
$\Phi_{2}$ is compactified.

\section{Proof of unitary equivalence}

In this appendix it will be demonstrated that
representations of the W-algebra on the
modules $\mathcal{W}^{j}_{mn}$ and $\mathcal{W}^{-j-1}_{mn}$
are unitarily equivalent. We restrict the discussion
to the holomorphic copy of the algebra. 

The Fock vacua $|jmn\rangle$ diagonalize the action of the
zero modes $\mathsf{W}_{s,0}$. Consider the eigenvalue equation
\be
\mathsf{W}_{s,0}|jmn\rangle=W_{s}(j,m,n)|jmn\rangle
\ee
To prove unitary equivalence of representations to spin $j$ and
spin $-j-1$ it suffices to show
\be
W_{s}(j,m,n)=W_{s}(-j-1,m,n)
\ee
A clarifying remark is necessary here. The norm of an arbitrary state in the module $\mathcal{W}^{j}_{mn}$ is
given by
\begin{multline}
||\mathsf{W}_{s_k,-n_{k}}(j,m,n)\dots \mathsf{W}_{s_1,-n_{1}}(j,m,n)\Omega||^2
=\\(\Omega,\mathsf{W}_{s_1,n_{1}}(j,m,n)\dots\mathsf{W}_{s_k,n_{k}}(j,m,n)\mathsf{W}_{s_k,-n_{k}}(j,m,n)\dots \mathsf{W}_{s_1,-n_{1}}(j,m,n)\Omega)
\end{multline}
In order to evaluate this expression, we have to move the
annihilation operators $\mathsf{W}_{s,n}(j,m,n)$ with $n>0$ to the right
until they hit the vacuum. Using the commutation relations of the
W-algebra, this produces zero-modes $\mathsf{W}_{s,0}(j,m,n)$ which upon
application of the zero-modes to the vacuum 
yields the corresponding eigenvalues $W_{s}(j,m,n)$. The norm is thus
invariant under $j\rightarrow -j-1$ if $W_{s}(j,m,n)=W_{s}(-j-1,m,n)$.
In order to prove the invariance of the eigenvalues under
$j\rightarrow -j-1$ we need to remember that  
the W-algebra currents $\mathsf{W}_{s}(z)$ are polynomials
in $\partial\phi_{1}(z),\partial\phi_{2}(z)$ and higher derivatives
thereof. We furthermore have to keep in mind the mode expansions
\be\label{modex}
\partial \phi_{1}(z)=-i \sum_{n}\mathsf{a}^{(1)}_{n}z^{-n-1}\, , \qquad
\partial \phi_{2}(z)=-i \sum_{n}\mathsf{a}^{(2)}_{n}z^{-n-1}
\ee
where the Euclidean momentum zero-modes $\mathsf{a}_{0}^{(k)}$ are related
to their Minkowskian counterparts through
\be
\mathsf{a}^{(1)}_{0}=\mathsf{p}_{1}+\frac{ib}{2}\, , \qquad\mathsf{a}^{(2)}_{0}=\mathsf{p}_{2}^{L}
\ee
From the structure of the OPE it follows that we can always normalize
the currents $\mathsf{W}_{s}(z)$ such that their modes
 match the requirement of hermiticity,
\be
\mathsf{W}_{s,n}^{\dagger}=\mathsf{W}_{s,-n}
\ee
This implies that the numerical coefficients of the above mentioned
 polynomials are real for $s$ even and imaginary for $s$ odd.
Furthermore, with the mode expansions (\ref{modex}) and the behaviour
of the currents under parity $\phi_{2}\rightarrow -\phi_{2}$,
\be\label{parity}
\mathsf{W}_{s}[\phi_{1},-\phi_{2}]=(-)^{s}\mathsf{W}_{s}[\phi_{1},\phi_{2}]
\ee
we claim that it follows that the zero-mode eigenvalues $W_{s}(j,m,n)$
may be represented as polynomials in $j$ according to
\be
W_{s}(j,m,n)=\sum_{k=0}^{s} C^{s}_{k}(m,n)j^{k}
\ee
with {\it real} coefficients $C^{s}_{k}(m,n)$. In order to prove this statement
consider a monomial of the form
\be\label{M}
\mathfrak{M}(z)=C:(\partial^{n_1}\phi_1)^{k_1}(\partial^{n_2}\phi_1)^{k_2}\times \dots
\times (\partial^{m_1}\phi_2)^{l_1}(\partial^{m_2}\phi_2)^{l_2}\dots :
\ee
which is a typical building block of the current $\mathsf{W}_{s}(z)$ given that
$\sum (n_i k_i+m_i l_i)=s$. The constant $C$ is real for $s$ even and imaginary for $s$ odd, as argued above. Now consider the action of the zero-mode
$\mathfrak{M}_0$ of $\mathfrak{M}(z)$ on the highest weight state $|jmn\rangle$. Taking into account
the mode expansions (\ref{modex}) we find
\begin{multline}
\mathfrak{M}_{0}|jmn\rangle=C \bigl ( (n_1-1)!(-)^{n_1}i \mathsf{a}_{0}^{(1)}\bigr )^{k_1}
\bigl ( (n_2-1)!(-)^{n_2}i \mathsf{a}_{0}^{(1)}\bigr )^{k_2}\times \dots\\
\times \bigl ( (m_1-1)!(-)^{m_1}i \mathsf{a}_{0}^{(2)}\bigr )^{l_1}
\bigl ( (m_2-1)!(-)^{m_2}i \mathsf{a}_{0}^{(2)}\bigr )^{l_2}\dots|jmn\rangle
\end{multline}
which can be rewritten according to, where $\mathrm{const.}\in \mathbb{R}$,
\be
\mathfrak{M}_{0}|jmn\rangle=C\times \mathrm{const.}\times i^{\sum k_i+\sum l_i} (\mathsf{a}_{0}^{(1)})^{\sum k_i}(\mathsf{a}_{0}^{(2)})^{\sum l_i}|jmn\rangle
\ee
Taking properly into account that the action of the zero-mode $\mathsf{a}_{0}^{(1)}$ on
the state $|jmn\rangle$ is given by $\mathsf{a}_{0}^{(1)}|jmn\rangle=-ibj|jmn\rangle$,
we finally find
\be
\mathfrak{M}_{0}|jmn\rangle=C\times \mathrm{const.'}\times i^{\sum l_i} (\mathsf{a}_{0}^{(2)})^{\sum l_{i}} j^{\sum k_i}|jmn\rangle
\ee
where $\mathrm{const.'}$ is real at least for real $b$ resp. $k>2$.
Now, $\sum l_{i}$ counts the number of occurrences of the field $\phi_{2}$ in $\mathfrak{M}$.
Therefore, due to (\ref{parity}), $\sum l_{i}$ is even for $s$ even and odd for $s$ odd
which implies that $i^{\sum l_i}$ is real for $s$ even and imaginary for $s$ odd.
However, the constant $C$ that appears in (\ref{M}) has the same property, as demonstrated
above. This means, since $\mathsf{a}_{0}^{(2)}$ is hermitean and has real eigenvalues,
that the coefficient of $j^{\sum k_i}$ in $\mathfrak{M}_{0}|jmn\rangle$ is real
which proves the asserted statement.

In order to conclude our proof of unitary equivalence of representations we note that
 hermiticity of $\mathsf{W}_{s,0}$ implies
\be
(W_{s}(j,m,n))^{\ast}=W_{s}(j,m,n)
\ee
However, in the principal continuous series we have $j\in -\frac{1}{2}+i\mathbb{R}$ and $j^{\ast}=-j-1$. It therefore follows from the reality
of the $W_{s}(j,m,n)$ and the reality of the coefficients
$C^{s}_{k}(m,n)$ that
\be
W_{s}(j,m,n)=W_{s}(-j-1,m,n)
\ee
and the proof is complete.

\section{Proof of braid relations}
Let us sketch only the basic ideas that enter the proof of (\ref{exchi}). It is first
of all useful to restrict the range of $\sigma_{1}$ and $\sigma_{2}$ to
the fundamental intervall $|\sigma_{1}-\sigma_{2}|<2\pi,\,\, \sigma_{1}<\sigma_{2}$.
Furthermore, we find it convenient to introduce a splitting of the screening
charge $\mathsf{Q}(\sigma)=\int \limits_{0}^{2\pi} d\varphi\, \mathsf{V}(\sigma+\varphi)$
according to
\be
\mathsf{Q}(\sigma_{1})&=&\mathsf{Q}_{I_{c}}+\mathsf{Q}_{I_{1}}\\
\mathsf{Q}(\sigma_{2})&=&\mathsf{Q}_{I_{c}}+\mathsf{Q}_{I_{2}}
\ee
where
\be\label{charges}
\mathsf{Q}_{I_{c}}=\int_{I_{c}} d\varphi\,\mathsf{V}(\varphi), \quad 
\mathsf{Q}_{I_{1}}=\int_{I_{1}} d\varphi\, \mathsf{V}(\varphi), \quad
\mathsf{Q}_{I_{2}}=\int_{I_{2}} d\varphi\, \mathsf{V}(\varphi)
\ee
with $I_{c}=[\sigma_{2},\sigma_{1}+2\pi],\,I_{1}=[\sigma_{1},\sigma_{2}],\, I_{2}=[\sigma_{1}+2\pi,\sigma_{2}+2\pi]$. We note that $\mathsf{Q}_{I_{1}}$ and $\mathsf{Q}_{I_{2}}$ are linearly related according to
\be
\mathsf{Q}_{I_{2}}=q^{4\mathsf{j}+4}\mathsf{Q}_{I_{1}}
\ee
which is a simple consequence of the monodromy property (\ref{monov}) of the screening 
current. One may then use the fact that
the screening current $\mathsf{V}$ obeys a simple exchange relation with
the building block $\mathsf{f}_{1}$ to derive that in the fundamental intervall
\be\label{Qg1}
\mathsf{Q}_{I_{c}}\mathsf{f}_{1}(\sigma_{1})=q\,\mathsf{f}_{1}(\sigma_{1})\mathsf{Q}_{I_{c}} \qquad
\mathsf{Q}_{I_{1}}\mathsf{f}_{1}(\sigma_{1})=q\,\mathsf{f}_{1}(\sigma_{1})\mathsf{Q}_{I_{1}} \qquad
\mathsf{Q}_{I_{2}}\mathsf{f}_{1}(\sigma_{1})=q^{3}\,\mathsf{f}_{1}(\sigma_{1})\mathsf{Q}_{I_{2}} \nonumber \\ \nonumber \\
\mathsf{Q}_{I_{c}}\mathsf{f}_{1}(\sigma_{2})=q\,\mathsf{f}_{1}(\sigma_{2})\mathsf{f}_{I_{c}} \qquad
\mathsf{Q}_{I_{1}}\mathsf{f}_{1}(\sigma_{2})=q^{-1}\,\mathsf{f}_{1}(\sigma_{2})\mathsf{Q}_{I_{1}} \qquad
\mathsf{Q}_{I_{2}}\mathsf{f}_{1}(\sigma_{2})=q\,\mathsf{f}_{1}(\sigma_{2})\mathsf{Q}_{I_{2}}
\ee\\
We may use these relations to demonstrate that there exist functions
$C^{ij}_{kl}(\mathsf{j})$ such that the following identities hold:
\be\label{gbraid12}
\mathsf{f}_{1}(\sigma_{1})\mathsf{f}_{2}(\sigma_{2})&=&\mathsf{f}_{2}(\sigma_{2})\mathsf{f}_{1}(\sigma_{1})\,C_{21}^{12}(\mathsf{j})+\mathsf{f}_{1}(\sigma_{2})\mathsf{f}_{2}(\sigma_{1})\,C_{12}^{12}(\mathsf{j})
\ee
\be\label{gbraid21}
\mathsf{f}_{2}(\sigma_{1})\mathsf{f}_{1}(\sigma_{2})&=&\mathsf{f}_{1}(\sigma_{2})\mathsf{f}_{2}(\sigma_{1})\,C_{12}^{21}(\mathsf{j})+\mathsf{f}_{2}(\sigma_{2})\mathsf{f}_{1}(\sigma_{1})\,C_{21}^{21}(\mathsf{j})
\ee\\
Finally, one may use the monodromy of the building blocks to generalize these
relations to arbitrary $\sigma_{1},\sigma_{2}$, with the result (\ref{exchi}).
I will sketch how this can be done.  Given arbitrary $\sigma_1,\sigma_2$, we
introduce the variable $\tilde{\sigma_2}$ according to
\be
\tilde{\sigma_2}=\sigma_2-\pi(\epsilon(\sigma_2-\sigma_1)-1)
\ee
One may convince oneself that this definition implies
$|\sigma_1-\tilde{\sigma_2}|<2\pi$ and $\sigma_1<\tilde{\sigma_2}$. Given that we want
to derive an exchange relation  for $\mathsf{f}_{1}(\sigma_1)\mathsf{f}_{2}(\sigma_2)$, we may reduce the calculation of the exchange relation
for $\mathsf{f}_{1}(\sigma_1)\mathsf{f}_{2}(\sigma_2)$ to the calculation
of the exchange relation for $\mathsf{f}_{1}(\sigma_1)\mathsf{f}_{2}(\tilde{\sigma_2})$. This is done by simply  setting $\sigma_2=\tilde{\sigma_2}+\pi(\epsilon(\sigma_2-\sigma_1)-1)$ and using the monodromy of the building blocks which
is given by
\be
\mathsf{f}_2 (x_{+}+2\pi)=e^{2\pi b(-\mathsf{p}_{1}+i\eta\mathsf{p}_2^L +\frac{ib}{2k})}\mathsf{f}_2 (x_{+})\\
\mathsf{f}_1 (x_{+}+2\pi)=e^{2\pi b(\mathsf{p}_{1}+i\eta\mathsf{p}_2^L +\frac{ib}{2k})}\mathsf{f}_1 (x_{+})
\ee
Now, since $|\sigma_1-\tilde{\sigma_2}|<2\pi$ and $\sigma_1<\tilde{\sigma_2}$,
we may apply the exchange relation (\ref{gbraid12}) and finally
by expressing $\tilde{\sigma_2}$ in terms of $\sigma_2$ and again
using the monodromy relations of the building blocks, we arrive at
an exchange relation for $\sigma_1,\sigma_2$ arbitrary.

It remains to consider to exchange relations satisfied by $\mathsf{f}_i (\sigma_1) \mathsf{f}_j (\sigma_2)$ for $i=j$. The exchange relations of the building block $\mathsf{f}_{1}$ resp.
$\bar{\mathsf{f}}_{1}$ with itself is a simple consequence
of (\ref{exe}) resp. the corresponding relation for the antichiral building blocks. To prove the relation (\ref{ff}) for the building block $\mathsf{f}_{2}$ (resp. $\bar{\mathsf{f}}_{2}$) is more delicate.
Here we may use the fact that the screening current  $\mathsf{V}(x_{+})$
obeys a simple exchange relation with itself according to
\be
\mathsf{V}(\sigma_{1})\mathsf{V}(\sigma_{2})=e^{-2\pi i b^{2}\epsilon(\sigma_{1}-\sigma_{2})}\mathsf{V}(\sigma_{2})\mathsf{V}(\sigma_{1})
\ee
This implies for the screening charges (\ref{charges}) , restricting
again to the fundamental interval $|\sigma_{1}-\sigma_{2}|<2\pi, \sigma_{1}<\sigma_{2}$, the following set of exchange relations:
\be\label{QQ}
\mathsf{Q}_{I_{1}}\mathsf{Q}_{I_{2}}=q^{4}\,\mathsf{Q}_{I_{2}}\mathsf{Q}_{I_{1}} \qquad \mathsf{Q}_{I_{c}}\mathsf{Q}_{I_{1}}=q^{-2}\,\mathsf{Q}_{I_{1}}\mathsf{Q}_{I_{c}} \qquad \mathsf{Q}_{I_{c}}\mathsf{Q}_{I_{2}}=q^{2}\,\mathsf{Q}_{I_{2}}\mathsf{Q}_{I_{c}}
\ee
One may use these relations and (\ref{Qg1}) to show that in the fundamental interval
\be
\mathsf{f}_{2}(\sigma_{1})\mathsf{f}_{2}(\sigma_{2})=q^{\frac{1}{k}}
\mathsf{f}_{2}(\sigma_{2})\mathsf{f}_{2}(\sigma_{1})
\ee
This can be generalized to arbitrary values of $\sigma_{1},\sigma_{2}$
by again using the monodromy properties of the building blocks.


\begin{thebibliography}{99}
\bibitem{witten}

E. Witten: {\it On string theory and black holes}, Phys. Rev. {\bf D44} (1991) 314.

\bibitem{dijkgraafverlinde}

R. Dijkgraaf, E. Verlinde, H. Verlinde: {\it String propagation in a black hole geometry}, Nucl. Phys. {\bf B371} (1992) 269.


\bibitem{bakas}

I. Bakas, E. Kiritsis: {\it Beyond the large $N$ limit: Non-linear $W_{\infty}$ as symmetry of the $SL(2,\mathbb{R})/U(1)$ coset model}, Int.J.Mod.Phys. {\bf A7} (1992) 55.







\bibitem{IntegrationQuantumPara}

C. Ford, G. Jorjadze, G. Weigt: {\it Integration of the $SL(2,\mathbb{R})/U(1)$ gauged WZNW theory by reduction and quantum parafermions}, Theor. Mat. Phys. {\bf 128} (2001) 1046; Teor. Mat. Fiz. {\bf 128} (2001) 249.

\bibitem{MuellerWeigtIntegrPeriodic}

U. M\"uller, G. Weigt: {\it Integration of the $SL(2,\mathbb{R})/U(1)$ gauged WZNW model with periodic boundary conditions}, Nucl. Phys. {\bf B568} (2000) 457.




\bibitem{MuellerComplSol}

U. M\"uller, G. Weigt: {\it The complete solution of the classical $SL(2,\mathbb{R})/U(1)$ gauged WZNW field theory}, Commun. Math. Phys. {\bf 205} (1999) 421.

\bibitem{AnalSolMueller}

U. M\"uller, G. Weigt: {\it Analytical solution of the $SL(2,\mathbb{R})/U(1)$ WZNW black hole model}, Phys. Lett. {\bf B400} (1997) 21.

\bibitem{LiouvilleRevisited}

J. Teschner: {\it Liouville theory revisited}, Class. Quant. Grav. {\bf 18} (2001) R153.

\bibitem{opexH3}

J. Teschner: {\it Operator product expansion and factorization in the $H_{3}^{+}$ WZNW model}, Phys. Lett. {\bf B571} (2000) 555.


\bibitem{crossymH3}

J. Teschner: {\it Crossing symmetry in the $H_{3}^{+}$ WZNW model}, Phys. Lett. {\bf B521} (2001) 127.

\bibitem{strucconst}

J. Teschner: {\it On structure constants and fusion rules in the $SL(2,\mathbb{C})/SU(2)$ WZNW model}, Nucl. Phys. {\bf B546} (1999) 390.
















\bibitem{dixon}
L.J. Dixon, M.E. Peskin, J. Lykken: {\it N=2 superconformal symmetry and $SO(2,1)$ current algebra}, Nucl. Phys. {\bf B325} (1989) 329.


\bibitem{nicolai}
Y. Kazama, H. Nicolai: {\it On the exact operator formalism of two-dimensional Liouville quantum gravity in Minkowski space-time}, Int.J.Mod.Phys. {\bf A9} (1994) 667.





\bibitem{zamo2}

A.B. Zamolodchikov, V.A. Fateev: {\it Nonlocal (parafermion) currents in two-dimensional conformal
quantum field theory and self-dual critical points in $Z_{N}$-symmetric statistical systems}, Sov.Phys. JETP {\bf 62} (1985) 215.























\bibitem{wz}

J. Wess, B. Zumino: {\it Consequences of anomalous Ward identities}, Phys. Lett. {\bf B37} (1971) 95.

\bibitem{novi}

S. P. Novikov: {\it The Hamiltonian formalism and a many valued analog of morse theory}, Usp. Math. Nauk. {\bf 37} (1982) 3.

\bibitem{witten2}

E. Witten: {\it Nonabelian Bosonization in two dimensions}, Comm. Math. Phys. {\bf 92} (1984) 455.

\bibitem{giveon1}

A. Giveon, D. Kutasov, O. Pelc: {\it Holography for Non-Critical Superstring}, JHEP {\bf 9910} (1999) 035.

\bibitem{giveon2}

A. Giveon, D. Kutasov: {\it Little String Theory in a Double Scaling Limit}, 
JHEP {\bf 9910} (1999) 034.

\bibitem{giveon3}

A. Giveon, D. Kutasov: {\it Comments on Double Scaled Little String Theory}, 
JHEP {\bf 0001} (2000) 023


\end{thebibliography}
\end{document}